\documentclass[twocolumn,showpacs,preprintnumbers,amsmath,amssymb]{revtex4-1}

\usepackage{graphicx}
\usepackage{dcolumn}
\usepackage{simplewick}
\usepackage{bm}
\usepackage{hyperref}
\usepackage[usenames,dvipsnames]{color}

\def\red{|\!|}

\begin{document}

\title{Triple excitation in relativistic coupled-cluster theory and properties
       of one-valence systems Rb and Sr$^+$}

\author{B. K. Mani and D. Angom}
\affiliation{Physical Research Laboratory,
             Navarangpura-380009, Gujarat,
             India}

\begin{abstract}
   We examine the contributions from triple excitation cluster operators in 
the relativistic coupled-cluster theory in atoms and ions. For this, we 
propose a tensor representation of the triple cluster operator. Based on this
representaion and using diagrammatic analysis, we  derive the linearized 
coupled-cluster equations for single, double and triple excitation cluster
operators. The contributions from the triple cluster operators to the hyper 
fine structure constants of single-valence systems Rb and Sr$^+$ are analysed  
using the perturbed triples. 
\end{abstract}



\maketitle


\section{Introduction}

   The coupled-cluster (CC) theory \cite{coester-58,coester-60,bartlett-07}
is an all-order many-body theory. It has proved to be one of the most reliable 
and accurate methods for precision atomic theory calculations. Apart from 
atomic systems \cite{nataraj-08,pal-07}, it has also been used with great 
success in nuclear \cite{hagen-08}, molecular \cite{isaev-04} and condensed 
matter \cite{bishop-09} calculations. In atoms and ions, calculations using
the relativistic coupled-cluster (RCC) theory has provided some of the best 
results. These include calculation of atomic electric dipole moments 
\cite{nataraj-08,latha-09}, hyperfine structure constants 
\cite{pal-07,sahoo-09}, electromagnetic transition properties 
\cite{thierfelder-09,sahoo-09a} and most importantly 
the NSI-PNC \cite{wansbeek-08}. 

  Among different flavours of CC theory, the coupled-cluster singles and 
doubles (CCSD) is a widely used approximation. However, for precision atomic 
calculations it is imperative to estimate the contributions from clusters of 
higher excitations. For CCSD approximation, the triple excitation cluster 
operators is the closest level of excitation neglected in the calculations. 
So, the leading order correction to CCSD is the effect of triple excitation
cluster operators. Further more, the triple excitations are expected to have
significant contributions in open shell systems. Due to the $N_v^3$ scaling,
where $N_v$ is the number of virtual orbitals, inclusion of triple excitation 
cluster operators pose severe computational challenges. A practical approach 
is selective inclusion of triple excitation cluster operators. Such 
calculations are important to make uncertainty estimates. In this work we 
examine contributions from the valence triple excitation cluster operators 
$S_3$ in RCC and propose a representation of $S_3$. Furthermore, to quantify 
the importance we carry out extensive calculations with different forms of 
perturbative triple excitations.

  Atomic parity non-conservation (PNC) is one class of atomic theory 
calculations, where precision theory calculations are important and 
uncertainty estimates are a must. The atomic theory results of PNC 
observable when combined with the experimental data provide estimates of 
parameters in standard model (SM) of particle physics \cite{dzuba-02}. Any 
deviation from the predictions of SM is an indication of new physics. 
The PNC in atoms occurs in two forms, nuclear spin-independent (NSI) and
nuclear spin-dependent (NSD). The former, has been theoretically and
experimentally studied in great detail, and the most accurate theoretical 
\cite{dzuba-02} and experimental \cite{wood-97} results are in the case atomic 
Cs. For the later, however, there are few theoretical studies. These are using 
many-body perturbation theory (MBPT) \cite{sahoo-11}, configuration 
interaction (CI) \cite{geetha-98,angom-99} and CI+MBPT 
\cite{dzuba-11a,dzuba-11b}.

 We recently proposed an RCC based method to incorporate nuclear 
spin-dependent interaction Hamiltonian as perturbation. The method is used
to calculate the NSD-PNC of Cs, Ba$^+$  and Ra$^+$ \cite{mani-11a} with 
associated {\em rms} uncertainties of 7\%, 4.4\% and 7.6\%, respectively. The 
details of the proposed theory are presented in another work of ours
\cite{mani-11b}. We believe that its possible to reduce the uncertainty, 
and the first step towards this could be the inclusion of triples cluster 
operators in RCC. 

  The paper is divided into nine sections. In Section. II, we give a brief 
review of CCSD. It is based on our previous works \cite{mani-09,mani-10} on
RCC of closed-shell and one-valence systems. Then the next section, 
Section. III, forms the core of the present work and describes the 
perturbative $S_3$. It discusses the possible chanels through which 
$S_3$ can arise and describes the tensor structure. In Section. IV, we give 
linearized RCC equations for singles, doubels and triples in terms of the CC 
excitation amplitudes. The HFS constants calculation using CCSD is breifly 
demonstrated in Section. V for the easy reference. A detailed description of 
HFS terms in RCC properties calculations and diagrams from the perturbative 
triples are given in the Sections. VI and VII. And in Section. VIII, we 
present and discuss our results.


\section{Brief review of RCC in CCSD approximation}

\begin{figure}[h]
  \includegraphics[width = 8cm]{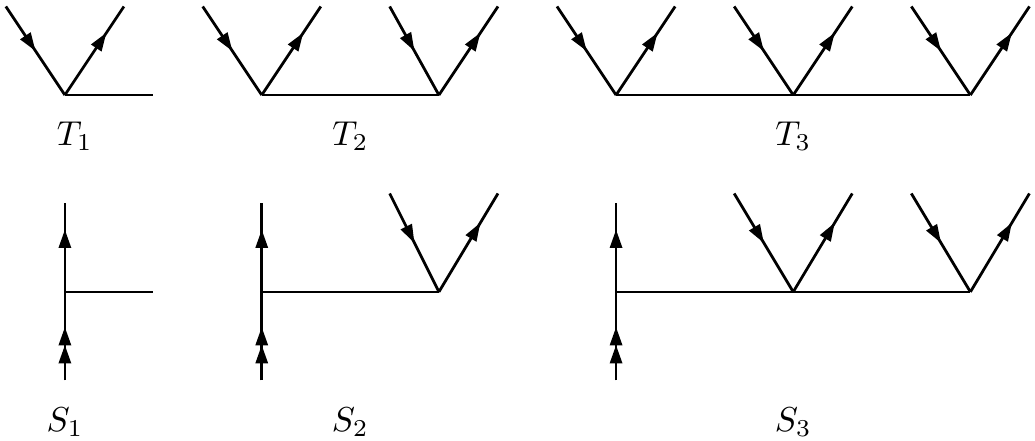}
  \caption{Diagrammatic representation of $T_1$ and $T_2$ operators. 
           The orbital lines with up (down) arrow indicate particle 
           (hole) states.}
  \label{fig_ts}
\end{figure}

    The Dirac-Coulomb Hamiltonian which accounts for the leading order
relativistic effects of atoms or ions with $N$ electrons is
\begin{equation}
  H^{\rm DC}=\sum_{i=1}^N\left [c\bm{\alpha}_i\cdot \mathbf{p}_i+
             (\beta_i-1)c^2 - V_N(r_i)\right ] +\sum_{i<j}\frac{1}{r_{ij}},
  \label{dchamil}
\end{equation}
where $\bm{\alpha}_i$ and $\beta$ are the Dirac matrices, $\mathbf{p}$ is the
linear momentum, $V_N(r)$ is the nuclear Coulomb potential and last term
is the electron-electron Coulomb interactions. For one-valence systems it
satisfies the eigen value equation
\begin{equation}
  H^{\rm DC}|\Psi_v\rangle = E_v|\Psi_v\rangle,
  \label{dcpsi}
\end{equation}
where $|\Psi_v\rangle$ and $E_v$ are the atomic state and energy,
respectively. In the CC method, the $|\Psi_v\rangle$ is expressed in terms 
of $T$  and $S$, the closed-shell and valence cluster operators 
respectively, as
\begin{equation}
  |\Psi_v\rangle = e^{T} \left [  1 + S \right ] |\Phi_v\rangle,
  \label{cceqn_1v}
\end{equation}
where $|\Phi_v\rangle$ is the one-valence Dirac-Fock reference state. It is
obtained by adding an electron to the closed-shell reference state,
$|\Phi_v \rangle = a^\dagger_v|\Phi_0\rangle$.  For an $N$ electron system,
which may be atom or ion, the cluster operators are
\begin{subequations}
\begin{eqnarray}
  T & = & \sum_{i=1}^{N-1} T_i,  \\
  S & = & \sum_{i = 1}^N S_i.
  \label{t}
\end{eqnarray}
\end{subequations}
The index $i$ represents the level of excitation ({\em loe}) of the 
cluster operators. Note that for $T$ {\em loe } is allowed up to the $N-1$ 
core electrons, where as for $S$ it is up to $N$ as it includes the valence 
electron. One major impediment to a full scale CC calculation is, the number 
of cluster operators proliferates exponentially with $i$ and calculations are 
unmanageable beyond the first few {\em loe}. This difficulty, as such, does not
diminish the applicability of CC. Most dominant correlation effects are 
incorporated in the first few {\em loe} and the approximation referred to
as the CC singles and doubles (CCSD) provides a very good description of 
the many-body effects. In this approximation
\begin{equation}
   T = T_1  + T_2, \text{ and }
   S = S_1 + S_2. 
\end{equation}
The operators in second quantization notation are
\begin{equation}
  T_1 = \sum_{a, p}t_a^p a_p^{\dagger}a_a, \text{ and }
  T_2 = \frac{1}{2!}\sum_{a b p q}t_{ab}^{pq}
  a_p^{\dagger}a_q^{\dagger}a_ba_a.
\end{equation}
\begin{equation}
  S_1 = \sum_{p}s_v^p a_p^{\dagger}a_v, \text{ and }
  S_2 = \sum_{a p q}s_{va}^{pq}
  a_p^{\dagger}a_q^{\dagger}a_aa_v.
\end{equation}
Here, $t_{\cdots}^{\cdots}$ and $s_{\cdots}^{\cdots}$ are the cluster
amplitudes. The indexes $abc\ldots$ ($pqr\ldots$) represent occupied (virtual)
states and $vwx\ldots$ represent valence states. The operators $T_1$ ($S_1$ )
and $T_2$ ($S_2$) give single and double replacements after operating on the
closed-(open-)shell reference states. 
The diagrammatic representations of $T$ and $S$ are shown in 
the Fig. \ref{fig_ts}.
%
%

    The closed-shell CC operators are the solutions of the nonlinear 
coupled equations
\begin{subequations}
\label{t_eqn}
\begin{eqnarray}
  \langle\Phi^p_a|\bar H_{\rm N}|\Phi_0\rangle = 0,
     \label{t1_eqn}                        \\
  \langle\Phi^{pq}_{ab}|\bar H_{\rm N}|\Phi_0\rangle = 0,
     \label{t2_eqn}
\end{eqnarray}
\end{subequations}
where $\bar H_{\rm N}=e^{-T}H_{\rm N}e^{T} $ is the
similarity transformed Hamiltonian and  the normal order Hamiltonian
$H_{\rm N} = H -\langle\Phi_0|H|\Phi_0\rangle$.  
The states $|\Phi^p_a\rangle$ and $|\Phi^{pq}_{ab}\rangle$ are the 
singly and doubly excited determinants, respectively. 
The details of the derivation are given in Ref. \cite{mani-09}.
The one-valence CC operators on the hand are obtained from the  
solutions of the equations
\begin{subequations}
\label{cc_sin_dou}
\begin{eqnarray}
  \langle \Phi_v^p|\bar H_N \! +\! \{\contraction[0.5ex]
  {\bar}{H}{_N}{S} \bar H_N S\} |\Phi_v\rangle
  &=&E_v^{\rm att}\langle\Phi_v^p|S_1|\Phi_v\rangle ,
  \label{ccsingles}     \\
  \langle \Phi_{va}^{pq}|\bar H_N +\{\contraction[0.5ex]
  {\bar}{H}{_N}{S}\bar H_N S\} |\Phi_v\rangle
  &=& E_v^{\rm att}\langle\Phi_{va}^{pq}|S_2|\Phi_v\rangle,
  \label{ccdoubles}
\end{eqnarray}
\end{subequations}
where $E_v^{\rm att} = E_v - E_0,$ is the attachment energy of the
valence electron. The details of the derivation of Eq. (\ref{cc_sin_dou})
we provide in our previous work \cite{mani-10}.


\section{Perturbative triples in RCC}

       Inclusion of perturbative triples to the CCSD approximation is 
refereed to as the CCSD(T) approximation. Within this approximation, 
Eq. (\ref{t}) is
\begin{equation}
   T = T_1  + T_2 + T_3 \text{ and }
   S = S_1 + S_2 + S_3.
\end{equation}
Here, $T_3$ and $S_3$ are the perturbative core and valence triple excitation 
cluster operators, respectively. Like the single and double operators, 
second quantized form of the triple excitation cluster operators are
\begin{subequations}
\begin{eqnarray}
  T_3 & = & \frac{1}{3!}\sum_{a b c p q r}t_{abc}^{pqr}
            a_p^{\dagger}a_q^{\dagger}a_r^{\dagger}a_ca_ba_a, \\
  S_3 & = & \frac{1}{2!}\sum_{a b p q r}s_{vab}^{pqr}
            a_p^{\dagger}a_q^{\dagger}a_r^{\dagger}a_ba_aa_v.
\end{eqnarray}
\end{subequations}
Previous calculations have shown contribution from $T_3$ to the properties of 
one-valence systems are much smaller than $S_3$, 
which is evident from the previous calculations. In particular the results 
reported in ref. \cite{sahoo-09}, where RCC is used. 
For this reason, in the present work, we consider and analyze in detail the
contribution from the $S_3$ cluster operators. The $S_3$ operator in RCC 
arise from two channels of contractions. First, residual Coulomb interaction
($V_2$) perturbs the  open shell operator $S_2 $, and second, $V_2$ 
perturbs the closed-shell operator $T_2$. 
Although, the contributions from the triples to properties are
small, it is imperative to include for high precision calculations. The
calculations related to discrete symmetry violations in atoms and ions belong 
to the class which require high precision. The need is even higher for 
open shell systems.


\subsection{Triples from $S_2$}

\begin{figure}[h]
\begin{center}
  \includegraphics[width = 8.5cm]{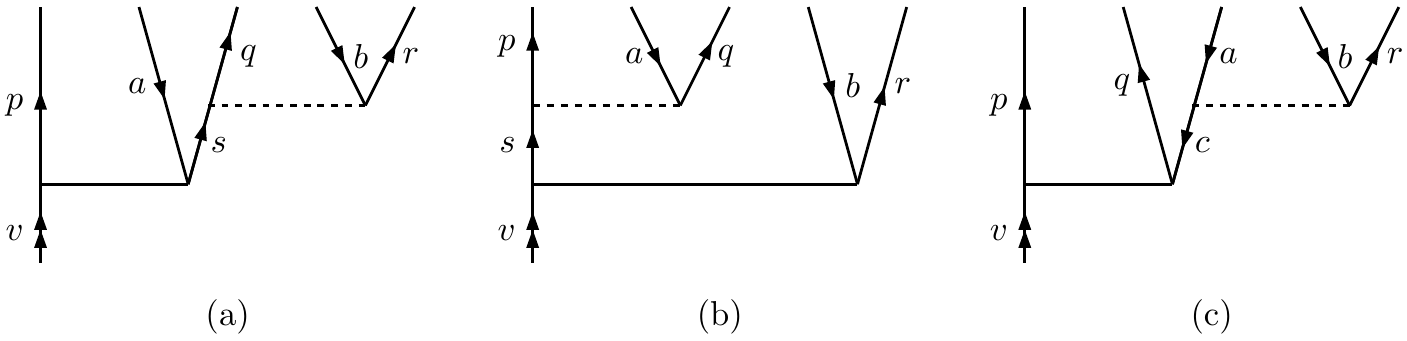}
  \caption{Perturbative triple cluster operators ($S_3$) diagrams 
           arising from the term $\contraction[0.5ex]{}{V}{_2}{S}V_2S_2$.
           The dashed line represents the Coulomb interaction.}
  \label{s3_from_s}
\end{center}
\end{figure}

  The Goldstone many-body diagrams of the first channel, $S_2$ perturbed 
triples, are shown in Fig. \ref{s3_from_s}. There are three topologically 
distinct diagrams. The first two arise from the contraction of $V_2$ with a 
virtual orbital of $S_2$ and the contribution is 
\begin{equation}
  s_{vab({\rm vs})}^{pqr}=\frac{1}{\epsilon _{vab}^{pqr}} 
     \sum_s \left (  v_{sb}^{qr}s_{va}^{ps} + v_{sa}^{pq}s_{vb}^{sr} \right ),
  \label{s3_vs}
\end{equation}
where, the energy denominator $\epsilon _{vab}^{pqr}= \epsilon_p + \epsilon_q  
+ \epsilon_r - \epsilon_v - \epsilon_a - \epsilon_b$, with $\epsilon$ as the 
orbital energies and the matrix elements in general are
$v_{ij}^{kl}=\langle kl|1/r_{12}|ij\rangle$, and 
$s_{ij}^{kl}=\langle kl|S_2|ij\rangle$. The subscripts 
$({\rm vs})$ indicate the cluster is from the contraction with
$S_2$ through a virtual orbital. In a similar way, the last diagram in 
%
%
Fig. \ref{s3_from_s} arises from the contraction of the core orbital and the
contribution is
\begin{equation}
  s_{vab({\rm cs})}^{pqr}=\frac{-1}{\epsilon _{vab}^{pqr}} 
     \sum_c   v_{cb}^{ar}s_{vc}^{pq}.
  \label{s3_cs}
\end{equation}
The negative sign follows from the application of Wick's theorem in the 
operator contractions. It is also evident from the rules of Goldstone diagram
evaluation \cite{lindgren-85}. According to which the phase of a diagram is
$(-1)^{l+h}$, where $l$ is the number of loops and $h$ is the number of 
internal core lines. For the present case, the diagram in 
Fig. \ref{s3_from_s}c has one internal core line and no loops. The subscript 
$(\rm{cs})$, like in previous case, indicate the origin of the 
term. Collecting the terms, the $S_2$ perturbed triples is
\begin{equation}
  s_{vab({\rm s})}^{pqr} = s_{vab({\rm vs})}^{pqr} + s_{vab({\rm cs})}^{pqr}.
\end{equation}
The two component $_{vab({\rm vs})}^{pqr}$ and $s_{vab({\rm cs})}^{pqr}$ have
different number of terms as $S_2$ is topologically asymmetric. Closed-shell
triples $T_3$, on the other hand, have one term each. 
\begin{figure}[h]
\begin{center}
  \includegraphics[width = 8.5cm]{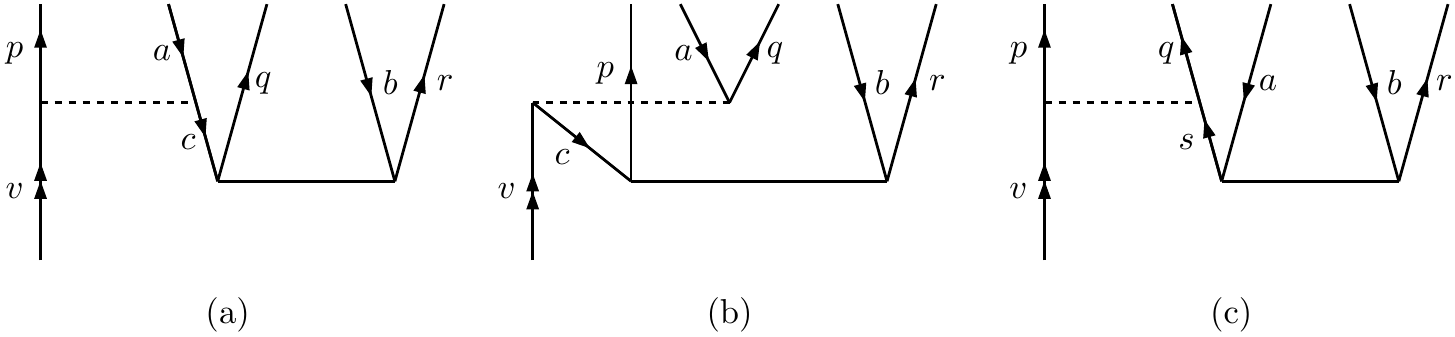}
  \caption{Perturbative triple cluster operators ($S_3$) diagrams 
           arising from the term $\contraction[0.5ex]{}{V}{_2}{T}V_2T_2$.
           The dashed line here is to represent the Coulomb interaction.}
  \label{s3_from_t}
\end{center}
\end{figure}
%
%


\subsection{Triples from $T_2$}

  Like in the $S_2$ perturbed triples, there are three Goldstone diagrams which 
contribute to the $T_2$ perturbed triples and these are shown in 
Fig. \ref{s3_from_t}. In the figure, the last diagram arises from the 
contraction of a virtual orbital and contribution is
\begin{equation}
  s_{vab({\rm vt})}^{pqr}=\frac{1}{\epsilon _{vab}^{pqr}} 
     \sum_c   v_{vs}^{pq}t_{ab}^{sr},
  \label{s3_vt}
\end{equation}
where, the subscript $({\rm vt})$ indicates perturbed $T_2 $ and contraction
of virtual orbital. The other two diagrams in the figure arise from the 
contraction of core orbital and contribution is 
\begin{equation}
  s_{vab({\rm ct})}^{pqr}=\frac{-1}{\epsilon _{vab}^{pqr}} 
     \sum_s \left (  v_{va}^{pc}t_{cb}^{qr} + v_{va}^{cq}t_{cb}^{pr} \right ).
  \label{s3_ct}
\end{equation}
The subscript $({\rm ct})$ indicates perturbed $T_2$ and contraction of core 
orbital. One key difference is noted when the above  expressions are 
compared with the $S_2$ perturbed triples. The number of diagrams arising from
the core and virtual orbital contractions are interchanged in the two cases. 
The $T_2$ perturbed triples is then
\begin{equation}
  s_{vab({\rm t})}^{pqr} = s_{vab({\rm vt})}^{pqr} + s_{vab({\rm ct})}^{pqr}.
\end{equation}
The total perturbed triples is the sum of the two contributions
\begin{equation}
    s_{vab}^{pqr} = s_{vab({\rm s})}^{pqr} + s_{vab({\rm t})}^{pqr}.
\end{equation}
All together, there are six perturbative $S_3$ diagrams. Three each from the
$S_2$ and $T_2$ perturbations. In contrast, for the perturbative $T_3$ there
are only two diagrams and one channel, $T_2$ perturbation.


\subsection{Tensor structure of $S_3$ }

  The diagrammatic form of triple excitation cluster operators as shown in 
Fig. \ref{fig_ts} are convenient representations. However, it runs into 
serious difficulties while decomposing into angular and radial parts. The 
central vertex, consisting of four lines, has no viable equivalent tensor
representation. To arrive at a consistent representation of the tensor
structure of the triple cluster operator, we analyze the angular reduction
of perturbative triple excitation diagrams shown in Fig. \ref{s3_from_s} and
\ref{s3_from_t}. As an example, we examine the perturbative  triples diagram 
in Fig. \ref{s3_from_s}(a).
%
%
\begin{figure}[h]
\begin{center}
  \includegraphics[width = 8.0cm]{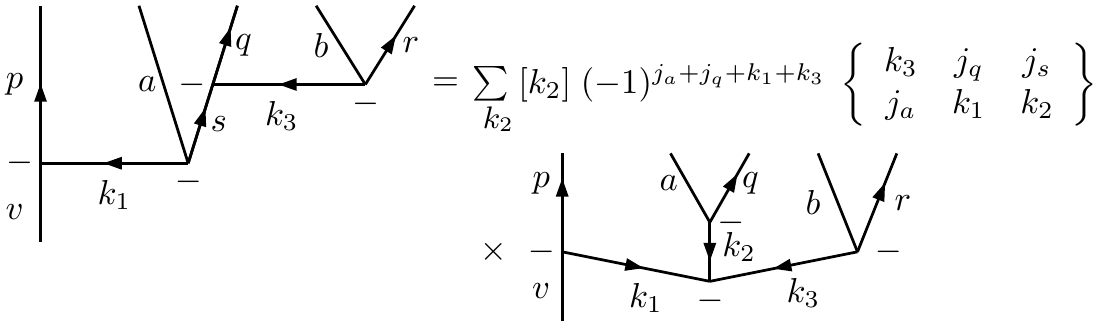}
  \caption{Angular reduction of the perturbative $S_3$ diagram shown in 
           Fig. \ref{s3_from_s}(a).}
  \label{triple_tensor}
\end{center}
\end{figure}
The angular reduction of the diagram into phase factor, 6$j$-symbol and an 
irreducible free angular diagram are shown in Fig. \ref{triple_tensor}. The
later represents the combinations of 3$j$-symbols to represent geometric part 
of the matrix element and remaining is the physical part, this follows from the 
Wigner-Eckert theorem. 

 There are other diagrammatic representations of the tensor structure 
of $S_3$. An example is the one given in Ref. \cite{porsev-06}, where an 
intermediate line is coupling of total angular momentum and tensor 
operator. In the present case, spin-orbitals are coupled pairwise to represent 
a matrix element of tensor operators of rank $k_i$ ($i=1, 2, 3$) which are 
again coupled. The tensor representation of $S_3 $ in explicit form is
\begin{eqnarray}
  S_3 &=& \sum_{k_1,k_2,k_3} s_3(k_1,k_2,k_3) \left \{ 
        \mathbf{C}_{k_1}(\hat{\mathbf{r}}_1) 
        \mathbf{C}_{k_2}(\hat{\mathbf{r}}_2) \right \}^{k_3} \nonumber \\
      &&\times \mathbf{C}_{k_3}(\hat{\mathbf{r}}_3),
\end{eqnarray}
where $\mathbf{C}_k(\hat{\mathbf{r}})$ are c-tensor operators of rank $k$,
and $\hat{\mathbf{r}}_i $ are unit vectors in the coordinates $\mathbf{r}_i$ 
of the $i$th electron. The notation $\{\ldots\}^{k_3}$ indicates coupling of
two c-tensors to a $k_3$ ranked c-tensor.   
The angular momenta and the rank of the tensor operators must
satisfy the triangular conditions
$|j_v - j_p|\leqslant k_1 \leqslant j_v + j_p$, 
$|j_a - j_q|\leqslant k_2 \leqslant j_a + j_q$, 
$|j_b - j_r|\leqslant k_3 \leqslant j_b + j_r$, and
$|k_1 - k_2|\leqslant k_3 \leqslant k_1 + k_2$. At the same time, 
spin-orbitals must satisfy the parity selection rule
$(-1)^{l_v + l_p}=(-1)^{l_a + l_q} = (-1)^{l_b + l_r}$. One important property
of the representation considered here is, the vertices in the tensor form of 
the triple operator can be inter changed with appropriate phase factor. In 
other words, there is an inherent symmetry in the coupling sequence 
considered.


\section{Triples from linearised RCC}

 Extending the CCSD approximation in RCC to include triple excitation is not 
a difficult proposition but entails enormous computational complications. In 
addition, there is several orders of magnitude increase in the number of 
cluster amplitudes. The diagrammatic analysis, albeit easier and tractable, is 
cumbersome as there is a large increase in the number of diagrams. An 
approximation, which incorporates the leading order effects of triple cluster 
amplitude but with much less computational complexity is the linearized 
treatment of the triples. The number of cluster amplitudes, however, are still 
large. For this we consider the inclusion of valence triples $S_3$ in the 
linearised RCC. From the definition of $\bar H_{\rm N}$, introduced in 
Eq. (\ref{t_eqn}), linearised RCC is equivalent to the approximations 
\begin{subequations}
  \label{hn_lin}
\begin{eqnarray}
  \bar H_{\rm N} &=& H_{\rm N} 
  + \contraction[0.5ex]{}{H}{_{\rm N}}{T} H_{\rm N} T,
                                             \\
  \contraction[0.5ex]{\bar}{H}{_{\rm N}}{S} \bar{H}_{\rm N} S &=& 
  \contraction[0.5ex]{}{H}{_{\rm N}}{S} H_{\rm N} S.    
\end{eqnarray}
\end{subequations}
To analyze the contributions from $S_3$, the valence cluster operator 
$S = S_1 + S_2 + S_3$, however, $T = T_1 + T_2$. The RCC equations of the
single and double excitation cluster amplitude, Eq. (\ref{cc_sin_dou}),
in linear approximation are
\begin{subequations}
\label{cc_sin_dou_lin}
\begin{eqnarray}
  \langle \Phi_v^p|H_{\rm N} 
  + \{\contraction[0.5ex]{}{H}{_N}{T} H_{\rm N} T\} 
  &+& \{\contraction[0.5ex]{}{H}{_N}{S} H_{\rm N} S\} |\Phi_v\rangle
                                              \nonumber \\
  &=&E_v^{\rm att}\langle\Phi_v^p|S_1|\Phi_v\rangle ,
  \label{ccsingles_lin}           \\
  \langle \Phi_{va}^{pq}|H_{\rm N} 
  + \{\contraction[0.5ex]{}{H}{_N}{T} H_{\rm N} T\}
  &+&\{\contraction[0.5ex]{}{H}{_N}{S} H_{\rm N} S\} |\Phi_v\rangle
                                              \nonumber \\
  &=& E_v^{\rm att}\langle\Phi_{va}^{pq}|S_2|\Phi_v\rangle.
  \label{ccdoubles_lin}    
\end{eqnarray}
\end{subequations}
Similarly, using the same definitions, the linearized equation of triple
excitation cluster operators is
\begin{eqnarray}
  \langle \Phi_{vbc}^{pqr}|\{\contraction[0.5ex]{}{H}{_N}{T} H_{\rm N} T\}
  +\{\contraction[0.5ex]{}{H}{_N}{S} H_{\rm N} S\} |\Phi_v\rangle
  = E_v^{\rm att}\langle\Phi_{vbc}^{pqr}|S_3|\Phi_v\rangle.
  \label{cctriples_lin}
\end{eqnarray}
The above equation of $S_3$, except for the absence of $H_{\rm N}$,  is very
similar to the single and double excitation cluster equations. This key 
difference is due to structure of $H_{\rm N}$,  which is either
one- or two-body operator. The triples excitation operator $S_3$ is, however, a 
three-body operator.

  A more illustrative way to write the RCC equations is to identify the
unique diagrams from the contractions and write the equivalent algebraic 
expressions. The linearized CC equation of singles, Eq. (\ref{ccsingles_lin}), 
in terms of the cluster amplitudes is then                            
\begin{eqnarray}                                     
   (\Delta E_v &-& \epsilon_v^p) s^p_v = \sum_{bq}\left( 
     \tilde{v}^{bp}_{qv} t^q_b + \sum_r\tilde{ v}^{bp}_{qr} s^{qr}_{bv}  
     - \sum_{c} v^{bc}_{qv} \tilde {t}^{qp}_{bc} \right )
                                \nonumber \\
     &+& \sum_{bcqr}\Big [ {\widetilde v}^{bc}_{qr} s^{pqr}_{vbc}
         + v^{bc}_{qr} \left (s^{qrp}_{vbc}- s^{qpr}_{vbc} \right ) \Big ].
\end{eqnarray} 
Here, $\Delta E_v = E_v^{\rm att} - \epsilon_v$ is the valence correlation 
energy of $|\Psi_v\rangle $,  $\epsilon_v^p = \epsilon_v - \epsilon_p$ and 
$\tilde {v}^{kl}_{ij} = v_{ij}^{kl} - v_{ji}^{kl} = v_{ij}^{kl}-v_{ij}^{lk}$,
is the antysymmetrized matrix element. Similarly, for compact notations
the antysymmetrized closed-shell and valence cluster amplitudes
are defined as $\tilde {t}^{kl}_{ij}$ and $\tilde {s}^{kl}_{ij}$, 
respectively. Like in $S_1$ equations, we retain the $T$ and $S$ from the 
CCSD equations in $S_2$ equation as well. However, from triples cluster 
amplitudes, we consider only the valence triples $S_3$. The 
Eq. (\ref{ccdoubles_lin}) in terms of cluster amplitudes is
\begin{eqnarray}                                         
  (\Delta E_v &-&  \epsilon_{vb}^{pq})s^{pq}_{vb} = v^{pq}_{vb} 
      + \sum_{r}\Big ( v^{pq}_{rb} s^r_v + v^{pq}_{vr} t^r_b \Big )
      -  \sum_{c} \tilde{v}^{cq}_{vb} t^p_c 
                                        \nonumber \\                    
  &+& \sum_{rc} \Big ( v^{pc}_{vr} \tilde {t}^{rq}_{cb}                      
      + v^{cq}_{rb} \tilde{s}^{pr}_{vc} - v^{pc}_{rb} s^{rq}_{vc} 
      -  v^{qc}_{rv} t^{rp}_{bc} -  v^{cp}_{vr} t^{rq}_{cb}
                                        \nonumber \\                    
  &-& v^{cq}_{br} s^{rp}_{cv} \Big ) + \sum_{rs} v^{pq}_{rs} s^{rs}_{vb} 
      + \sum_{cd} v^{cd}_{vb} t^{pq}_{cd} 
      + \sum_{cdr} v^{cd}_{rb} \Big ( s^{prq}_{vdc} 
                                        \nonumber \\
  &+& s^{rpq}_{vcd} - s^{prq}_{vcd}\Big ) 
      + \sum_{src} \Big ( \tilde {v}^{pc}_{sr} s^{srq}_{vcb}
      - v^{pc}_{sr} s^{srq}_{vbc} \Big )
                                        \nonumber \\
  &+& \left ( \begin{array}{c}                                                
            p\leftrightarrow q \\                                           
            v\leftrightarrow b                                              
           \end{array} \right ).                                             
  \label{lin_s2}                                                            
\end{eqnarray}                                                              
Here, $\epsilon_{vb}^{pq} = \epsilon_v + \epsilon_b - \epsilon_p
-\epsilon_q $ and $\bigl( \begin{smallmatrix}p\leftrightarrow q \\ 
v\leftrightarrow b \end{smallmatrix} \bigr ) $ indicates the combined 
permutations $p\leftrightarrow q$ and $v \leftrightarrow b$ of the previous 
terms within parenthesis. Interestingly, in this case terms with the combined
permutations represent topologically distinct diagrams. For $S_3$, the 
equation in terms of cluster amplitudes is
\begin{eqnarray}                                                            
  ( \Delta E_v &+& \epsilon_{vbc}^{pqr}) s^{pqr}_{vbc} =                  
     \left [ \sum_{s} \Big ( v^{qr}_{sc} s^{ps}_{vb}   
       +  v^{pr}_{sc} s^{sq}_{vb} + v^{pq}_{vs} t^{sr}_{bc} \Big )
                                             \right. \nonumber \\
   &-& \sum_{d} \Big ( v^{dr}_{bc} s^{pq}_{vd}
       - v^{dr}_{vc} t^{pq}_{db} - v^{pd}_{vb} t^{qr}_{dc} \Big )
       - \sum_{ds} \Big ( v^{pd}_{sb} s^{sqr}_{vdc}  
                        \nonumber \\                       
   &-&   v^{dr}_{bs} s^{pqs}_{vdc} - v^{rd}_{sc} s^{pqs}_{vbd} 
       - v^{dr}_{sc} s^{pqs}_{vbd} - v^{qd}_{bs} s^{spr}_{vdc}
                         \nonumber \\           
   &-& \left . v^{rd}_{cs} s^{pqs}_{vdb} \Big ) 
       + \sum_{su} v^{pq}_{su} s^{sur}_{vbc} \right ]
       + \left ( \begin{array}{c}
                q \leftrightarrow r \\ 
                b \leftrightarrow c
            \end{array} \right )
                         \nonumber \\ 
   &+& \sum_{su} v^{qr}_{us} s^{pus}_{vbc} 
       + \sum_{de} v^{de}_{bc} s^{pqr}_{vde}.
  \label{lin_s3}                                                            
\end{eqnarray}                                                              
Here, as defined earlier in the description of perturbed $S_3$, 
$\epsilon_{vbc}^{pqr}= \epsilon_v + \epsilon_b + \epsilon_c 
- \epsilon_p - \epsilon_q - \epsilon_r $. In the present case, the 
combined permutations $\bigl( \begin{smallmatrix}q\leftrightarrow r \\ 
b\leftrightarrow c \end{smallmatrix} \bigr )$ are just that, interchange
of the orbital lines and do not represent unique diagrams. Reason is, the
two permutations $q\leftrightarrow r$ and $b \leftrightarrow c$ are between 
orbitals of the same kind virtual and core, respectively. Where as in 
$S_2$ equations one of the permutations is between core and valence, which 
have different topological representations.


\section{HFS constants from RCC}

  The hyperfine interactions $H_{\rm hfs}$ are the coupling between nuclear 
electromagnetic moments and electromagnetic fields of atomic electrons. 
The interaction energies from $H_{\rm hfs}$ are the leading order corrections
to the atomic and ionic energies obtained from $H^{\rm DC}$. In terms of the 
tensor operators, the hyperfine interaction Hamiltonian is
\cite{johnson-07,charles-55}
\begin{equation}
  H_{\rm hfs} = \sum_i\sum_{k, q}(-1)^q t^k_q(\hat {\bf r}_i) T^k_{-q},
  \label{hfs_ham}
\end{equation}
where $t^k_q(\mathbf{r})$ and $T^k_{q}$ are irreducible tensor operators of rank
$k$ in the electron and nuclear spaces respectively.  For $k=1$, following 
parity selection rules, the allowed interaction is the magnetic dipole.
The explicit form of the associated tensor operators are
\begin{subequations}
\begin{eqnarray}
           T^1_q & = & \mu_q,  \\
  t^1_q({\bf r}) & = & \frac{-i\sqrt{2}[{\bm \alpha}\cdot{\bf C}_1
                        (\hat {\bf r})]_q} {cr^2},
  \label{hfs_magnetic}
\end{eqnarray}
\end{subequations}
where, ${\bf C}_1(\hat {\bf r})$ is a rank one tensor operator in electron
space and $\mu _q$ is a component of $\bm \mu$, the nuclear magnetic moment
operator. Interactions of higher rank multipoles are defined with similar 
form of tensor operators. However, these are not discussed as in this work
as we examine the  corrections to magnetic dipole hyperfine constants from 
the triples.
From the expression in Eq. \ref{hfs_ham}, we can write the magnetic dipole HFS 
constant as
\begin{equation}
   A = \frac{\langle \Psi_v|\sum_i\sum_q(-1)^q 
             t^1_q(\hat {\bf r}_i) T^1_{-q}|\Psi_v \rangle}
            {\langle \Psi_v |\Psi_v \rangle},
  \label{a_expect}
\end{equation}
where $q = -1, 0, 1$. The matrix element is calculated from the single 
particle reduced matrix element 
\begin{equation}
  a = \frac{g_I\mu_N}{\sqrt{j_v(j_v+1)(2j_v+1)}}
      \langle n_v\kappa_v ||t^1||n_v\kappa_v \rangle.
      \label{hfs_mdipole}
\end{equation}
Here, $g_I$ $(\mu = g_II\mu_N)$ is the gyromagnetic ratio, $\mu_N$ is the 
nuclear magneton and $|n_v\kappa_v \rangle $ is the valence single particle 
wave function. In a similar way, the HFS constants of higher order moments
may be calculated.

Using CC wave function from Eq. (\ref{cceqn_1v})
\begin{eqnarray}
  \langle \Psi_v|H_{\rm hfs}|\Psi_v \rangle &=& \langle \Phi_v|
     e^{T^\dagger}(1+ S)^\dagger H_{\rm hfs} e^T(1 + S)|\Phi_v\rangle,
                  \;\;\;\;\;\;     \nonumber \\
  &=&\langle \Phi_v| \tilde H_{\rm hfs} +  2 S^\dagger \tilde H_{\rm hfs} 
      + S^\dagger \tilde H_{\rm hfs} S |\Phi_v\rangle.
  \label{hfs_cc}
\end{eqnarray}
Where, $\tilde H_{\rm hfs}, = e^{T^\dagger} H_{\rm hfs} e^T,$ is the dressed
hyperfine interaction and it is a non terminating series of closed-shell CC 
operator $T$. Further more, 
$S^\dagger \tilde H_{\rm hfs}  = \tilde H_{\rm hfs} S $ is considered while
writing the equation. The higher order terms beyond second-order are, however, 
negligible and a truncated expression is considered. The approximation
\begin{equation}
  \tilde H_{\rm hfs} \approx H_{\rm hfs} + H_{\rm hfs} T + 
  T^\dagger H_{\rm hfs} + T^\dagger H_{\rm hfs} T,
  \label{hfs_truncate}
\end{equation}
accounts for all the important correlation effects and used in the present 
work. The normalization factor, denominator in Eq. (\ref{a_expect}), is 
\begin{equation}
  \langle \Psi_v |\Psi_v \rangle =
  \langle \Phi_v|\left (1 + S^\dagger\right ) e^{T^\dagger} e^T
  \left ( 1 +  S\right )|\Phi_v\rangle.
\end{equation}
Like in the dressed properties operator, $e{^T}^\dagger e^T$ is a 
non-terminating series. However, it is sufficient and accurate to 
consider up to the second order
\begin{eqnarray}
  \langle \Psi_v |\Psi_v \rangle & \approx & 
    \langle \Phi_v|\left (1 + S^\dagger S +   T^\dagger T + 
    S^\dagger T   
+ T^\dagger S \right )|\Phi_v\rangle.  \;\;\;\;\;\;
\end{eqnarray}
The last two terms, although finite, are expected to be small as the 
contribution is of the form $S_2^\dagger T_1$ and $T_1^\dagger S_2$, 
respectively. For this reason, these two terms are not included in our
calculations. 
%
%
\begin{figure}[h]
\begin{center}
  \includegraphics[width = 8.0cm]{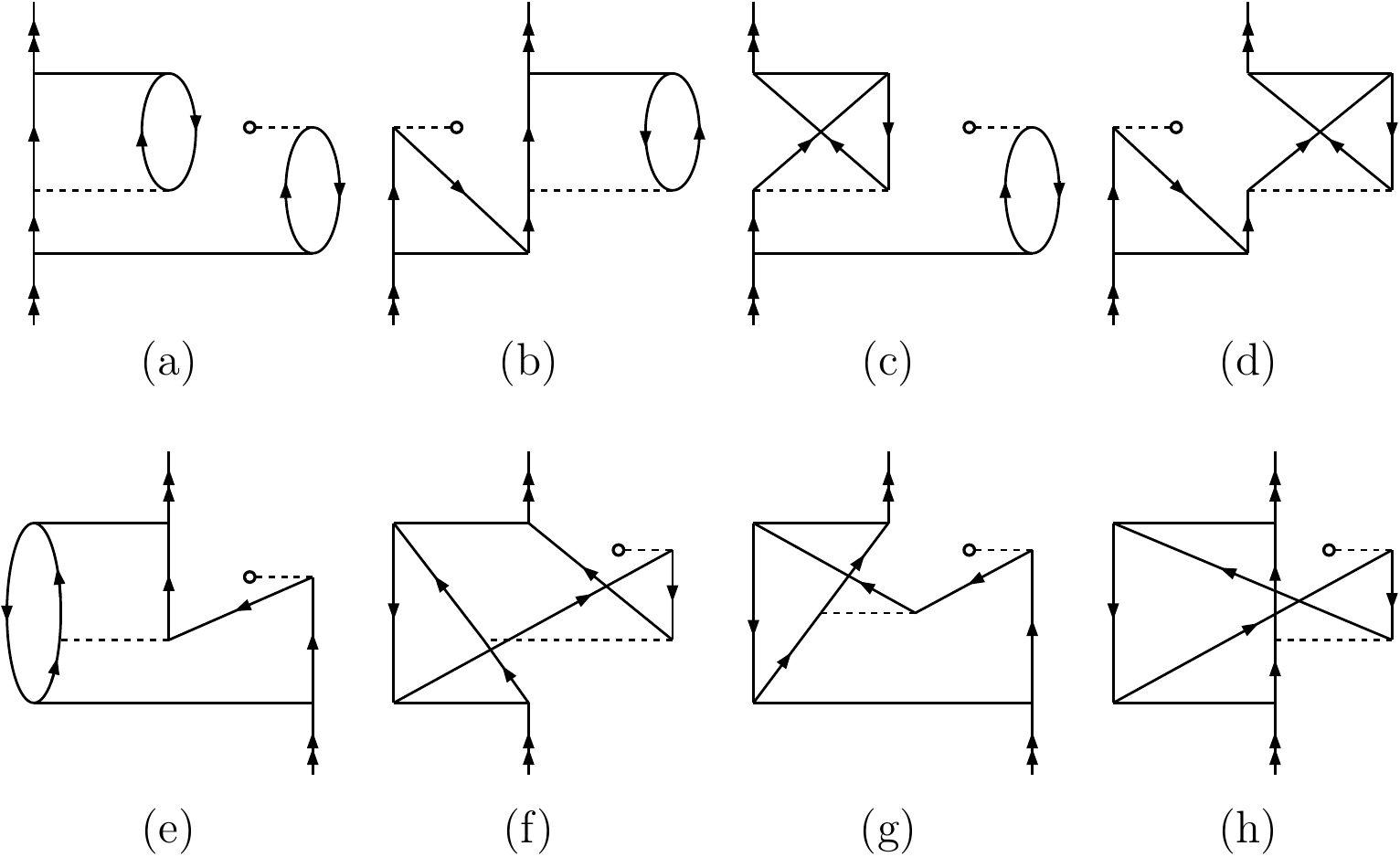}
  \caption{Goldstone HFS diagrams from 
           ${S_2}^\dagger h_{\rm hfs}S_{3({\rm vs})}$, which contribute to
           $A_{({\rm vs})}^1 $. In the first diagram (a) all the two body
           vertices, $S_2^\dagger$, $1/r_{12}$ and $S_2$ are of direct type.
           Remaining diagrams are combinations of exchange at different 
           vertices.}
  \label{hfs_vs12}
\end{center}
\end{figure}
%
%


\section{HFS constants from $S_2 $ perturbed triples}

  From the expression of HFS constant with RCC wave function in 
Eq. (\ref{hfs_cc}), the lowest order triples contributions are of the
form
\begin{equation}
  A_3 = T_1^\dagger H_{\rm hfs}S_3 + T_2^\dagger H_{\rm hfs}S_3 
        + S_2^\dagger H_{\rm hfs}S_3.
\end{equation}
The first term is, however, neglected in the present calculations. The reason 
is, the $T_1$ cluster amplitudes are small and have no significant 
contributions. For easy book keeping, contributions from the remaining two
terms is bifurcated based on the nature of $H_{\rm hfs}$ matrix elements. Two
of the possibilities, $\langle a|h_{\rm hfs}|p\rangle$ and 
$\langle v|h_{\rm hfs}|p\rangle$ are considered. There are 52 Goldstone 
HFS diagrams associated with these two matrix elements and the $S_2$ perturbed 
$S_3$. These are separated into groups and discussed in this section. The 
other forms, $\langle a|h_{\rm hfs}|a\rangle$ and  
$\langle p|h_{\rm hfs}|p\rangle$, enter through the structural radiation 
diagrams, which are negligibly small and are excluded from the present 
calculations. 
\begin{figure}[h]
\begin{center}
  \includegraphics[width = 8.0cm]{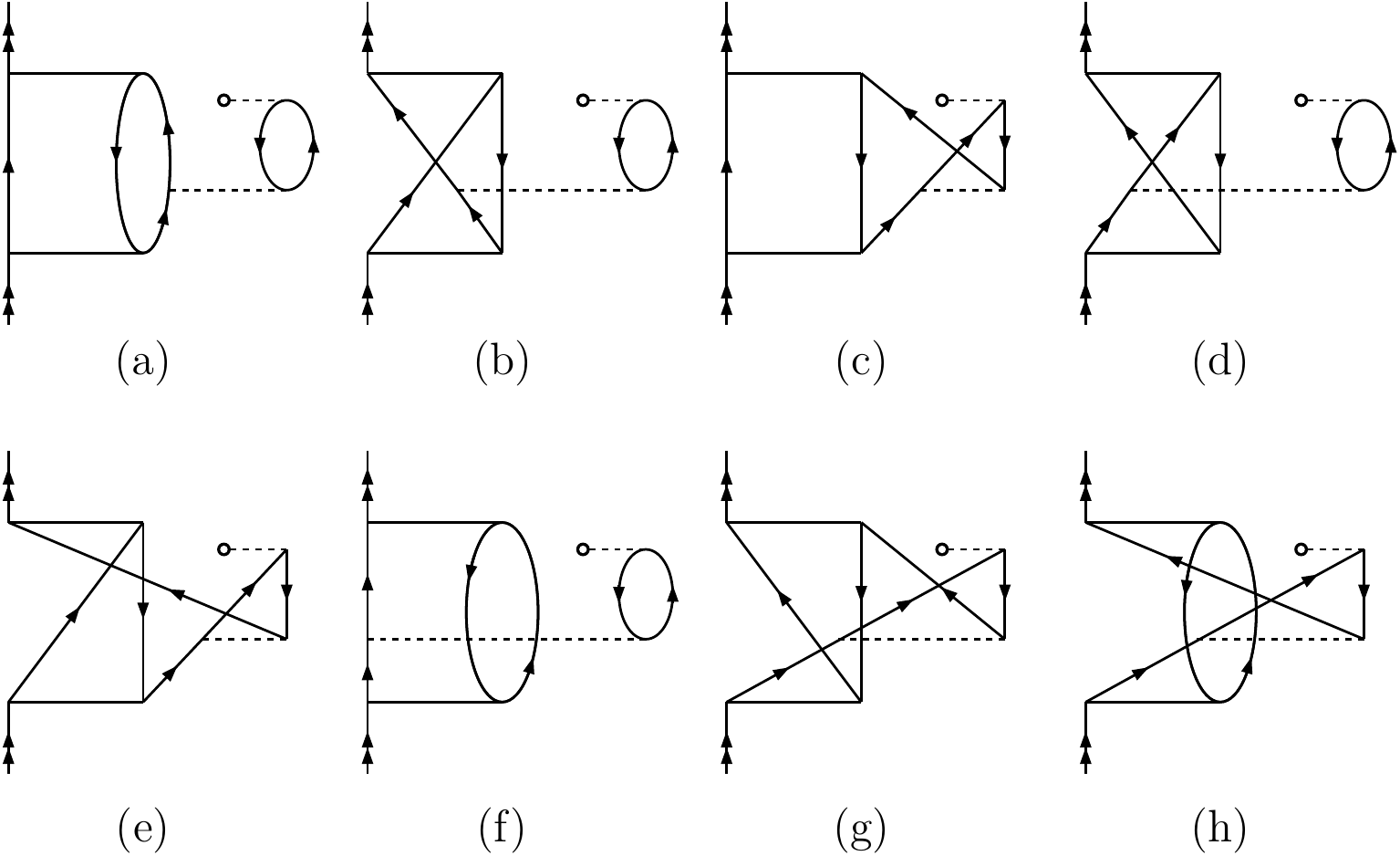}
  \caption{Goldstone HFS diagrams from 
           ${S_2}^\dagger h_{\rm hfs}S_{3({\rm vs})}$, which contribute to
           $A_{({\rm vs})}^2 $. In the first diagram (a) all the two body
           vertices, $S_2^\dagger$, $1/r_{12}$ and $S_2$ are of direct type.
           Remaining diagrams are combinations of exchange at different 
           vertices.  }
  \label{hfs_vs3}
\end{center}
\end{figure}
%
%


\subsection{Contribution from $S_2^\dagger H_{\rm hfs}S_{3({\rm vs})}$}

 Consider the triples of the form $s_{vab({\rm vs})}^{pqr}$ defined in 
Eq. (\ref{s3_vs}), for easy reference $S_{3({\rm vs})}$ define the general
form of the triples in this group. The contraction 
$\contraction[0.5ex]{}{h}{_{\rm hfs}}{S}h_{\rm hfs}S_{3({\rm vs})}$ is 
diagrammatically realized through four unique topologies. First, take the 
case where $h_{\rm hfs}$ has a core-particle contraction with $S_2$ in
$S_{3({\rm vs})}$ and contribution to $A$ is
\begin{eqnarray}
   A_{({\rm vs})}^1 & = &\sum_{vab}\sum_{pqrs} \frac{1}{\epsilon_{vab}^{pqr}}
               \left ( {s_{va}^{pq}}^* h_b^rv_{sa}^{pq}s_{vb}^{sr} 
              - {s_{va}^{pq}}^* h_b^rv_{sa}^{pq}s_{vb}^{rs}\right .
                    \nonumber \\
           && \left . - {s_{va}^{pq}}^* h_b^rv_{sa}^{qp}s_{vb}^{sr} 
              + {s_{va}^{pq}}^* h_b^rv_{sa}^{qp}s_{vb}^{rs} \right ),
\end{eqnarray}
where $h_b^r$ denotes the matrix element $\langle r|h_{\rm hfs}|b\rangle$.
The many-body diagrams in Fig. \ref{hfs_vs12}a-d are the representation of 
the above terms.  For compact notation, introduce the antisymmetrised 
representation of the residual Coulomb matrix element, 
$\tilde{v}_{ab}^{pq} = v_{ab}^{pq} - v_{ab}^{qp} = v_{ab}^{pq} - v_{ba}^{pq}$.
Antisymmetrised representation of the $s_{va}^{pq}$ is defined in the same
way.  In a more compact form
\begin{equation}
   A_{({\rm vs})}^1  = \sum_{vab}\sum_{pqrs} \frac{1}{\epsilon_{vab}^{pqr}}
      {s_{va}^{pq}}^* h_b^r\tilde{v}_{sa}^{pq}\tilde{s}_{vb}^{sr} .
\end{equation}
It must be noted that, the antisymmetrised form is employed for compact
notations. Otherwise, all the calculations are in non symmetrised 
representations and  is a better choice with diagrammatic analysis.

 Second, the core and particle lines of $h_{\rm hfs}$ contracts with the 
residual Coulomb and $S_2$, respectively. Diagrams arising from the 
contractions are shown in Fig. \ref{hfs_vs12}e-h and contribution is
\begin{eqnarray}
   A_{({\rm vs})}^2 & = &\sum_{vab}\sum_{pqrs} \frac{-1}{\epsilon_{vab}^{pqr}}
               \left ( {s_{bv}^{rq}}^* h_a^pv_{sa}^{rq}s_{bv}^{sp} 
               -{s_{bv}^{rq}}^* h_a^pv_{sa}^{rq}s_{bv}^{ps}  \right .
                    \nonumber \\
           && \left . -{s_{bv}^{rq}}^* h_a^pv_{sa}^{qr}s_{bv}^{sp} 
              +{s_{bv}^{rq}}^* h_a^pv_{sa}^{qr}s_{bv}^{ps} \right ).
\end{eqnarray}
In antisymmetrised representation
\begin{equation}
   A_{({\rm vs})}^2  = \sum_{vab}\sum_{pqrs} \frac{-1}{\epsilon_{vab}^{pqr}}
               {s_{bv}^{rq}}^* h_a^p\tilde{v}_{sa}^{rq}\tilde{s}_{bv}^{sp} .
\end{equation}
The two cases discussed so far have double virtual orbital contraction of 
$S_2^\dagger$ with either residual Coulomb interaction or $S_2$. As  a result
no unique diagrams arise from the anti-symmetrization of the $S_2^\dagger$. 
\begin{figure}[h]
\begin{center}
  \includegraphics[width = 8.0cm]{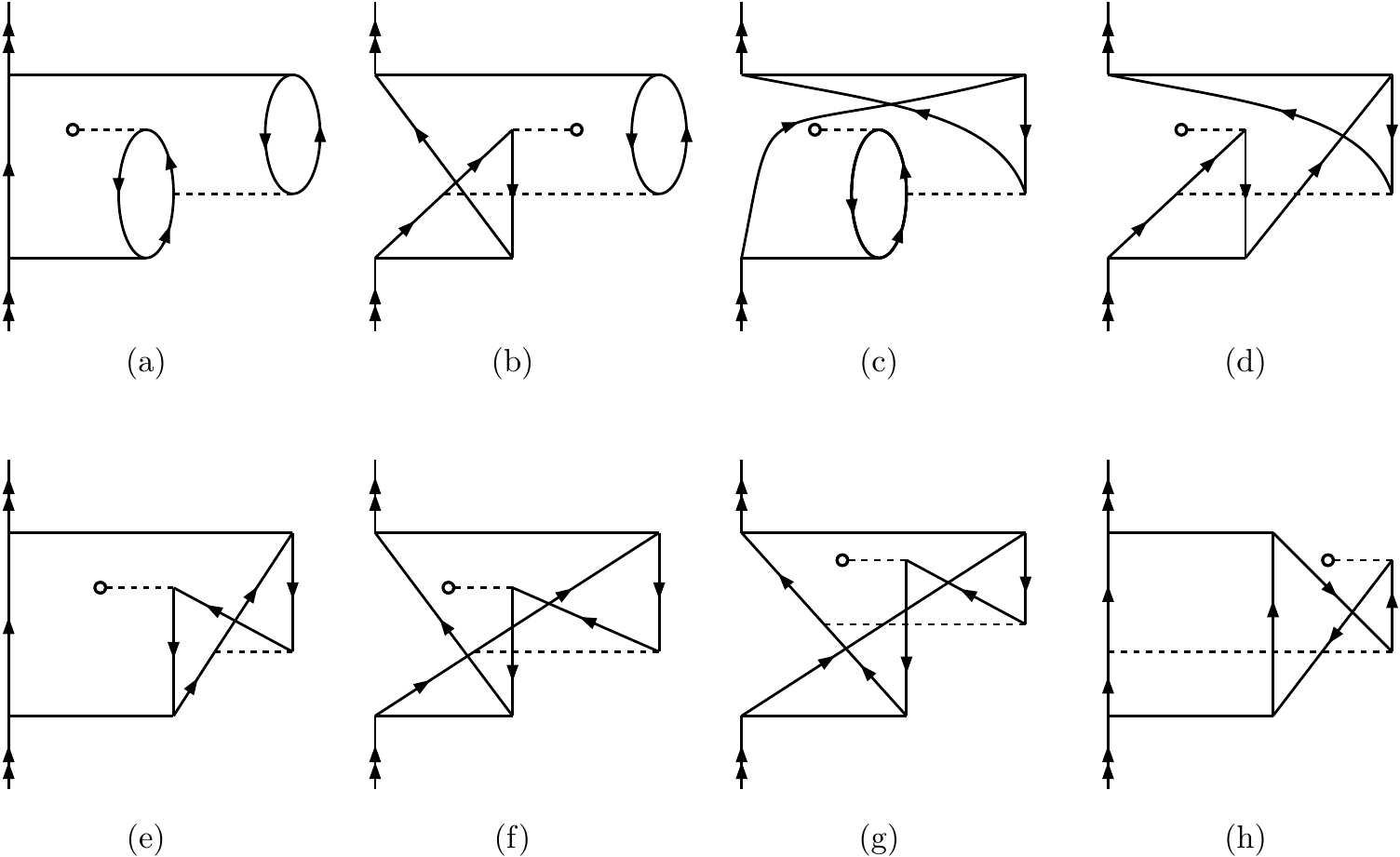}
  \caption{Goldstone HFS diagrams from 
           ${S_2}^\dagger h_{\rm hfs}S_{3({\rm vs})}$, which contribute to
           $A_{({\rm vs})}^3 $. In the first diagram (a) all the two body
           vertices, $S_2^\dagger$, $1/r_{12}$ and $S_2$ are of direct type.
           Remaining diagrams are combinations of exchange at different 
           vertices.  }
  \label{hfs_vs4}
\end{center}
\end{figure}

  Third, the core-virtual orbital lines of $h_{\rm hfs}$ contracts with the 
residual Coulomb interaction. Eight unique diagrams arise from the 
contractions and are shown in Fig.\ref{hfs_vs3}. The contribution is
\begin{equation}
   A_{({\rm vs})}^3  = \sum_{vab}\sum_{pqrs} \frac{1}{\epsilon_{vab}^{pqr}}
             \left [ {s_{va}^{pq}}^*h_r^b\tilde{v}_{sb}^{qr}\tilde{s}_{va}^{ps} 
             - {s_{va}^{qp}}^*h_r^b\tilde{v}_{sb}^{qr}\tilde{s}_{va}^{ps} 
             \right ],
 \label{a_vs3}
\end{equation}
here, as in previous  expressions the antisymmetrised $v$ and $S_2$ are used
for compact notations. The antisymmetrised expression of $S_2^\dagger$ can be
used to obtain the expression 
\begin{equation}
   A_{({\rm vs})}^3  = \sum_{vab}\sum_{pqrs} \frac{1}{\epsilon_{vab}^{pqr}}
       \tilde{s}_{va}^{pq*} h_r^b\tilde{v}_{sb}^{qr}\tilde{s}_{va}^{ps} . 
\end{equation}
There is a prominent difference of the present case from the previous two,
the exchange of $S_2^\dagger$ gives topologically unique diagrams. 

Finaly, the core and virtual orbital line of $h_{\rm hfs} $ contract with the
$S_2$ and residual Coulomb interaction, respectively. Diagrams from the 
contractions are shown in Fig. \ref{hfs_vs3}  and in antisymmetrised notations,
the contribution is 
\begin{equation}
   A_{({\rm vs})}^4  = \sum_{vab}\sum_{pqrs} \frac{1}{\epsilon_{vab}^{pqr}}
       \tilde{s}_{vb}^{pr*}h_q^b\tilde{v}_{sb}^{qr}\tilde{s}_{va}^{ps} . 
\end{equation}
In this case too, there are eight unique diagrams. 
\begin{figure}[h]
\begin{center}
  \includegraphics[width = 8.0cm]{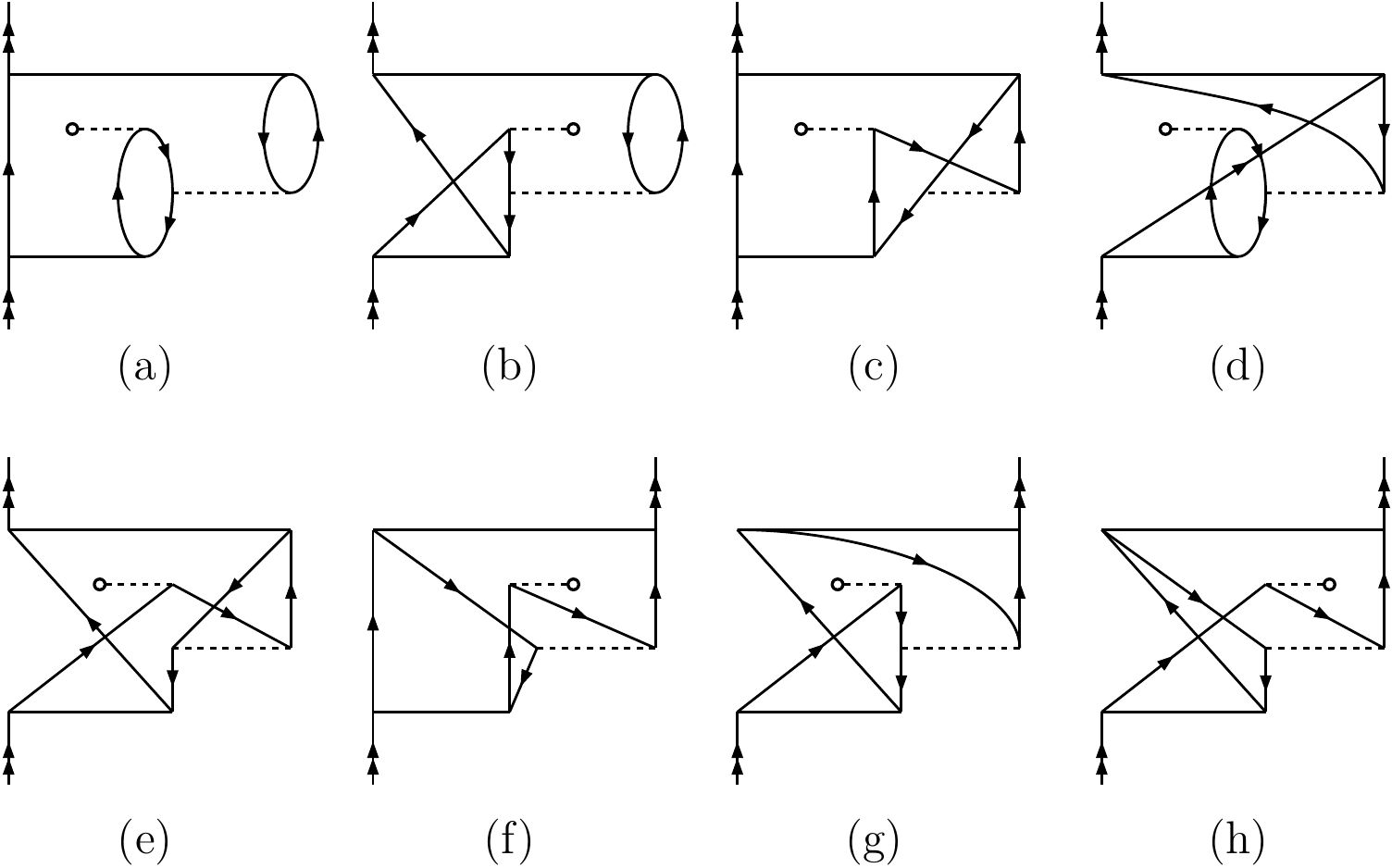}
  \caption{Goldstone HFS diagrams from 
           ${S_2}^\dagger h_{\rm hfs}S_{3({\rm cs})}$, which contribute to
           $A_{({\rm cs})}^1 $. In the first diagram (a) all the two body
           vertices, $S_2^\dagger$, $1/r_{12}$ and $S_2$ are of direct type.
           Remaining diagrams are combinations of exchange at different 
           vertices. }
  \label{hfs_cs1}
\end{center}
\end{figure}
%
%


\subsection{Contribution from $S_2^\dagger H_{\rm hfs}S_{3({\rm cs})}$}

\begin{figure}[h]
\begin{center}
  \includegraphics[width = 8.0cm]{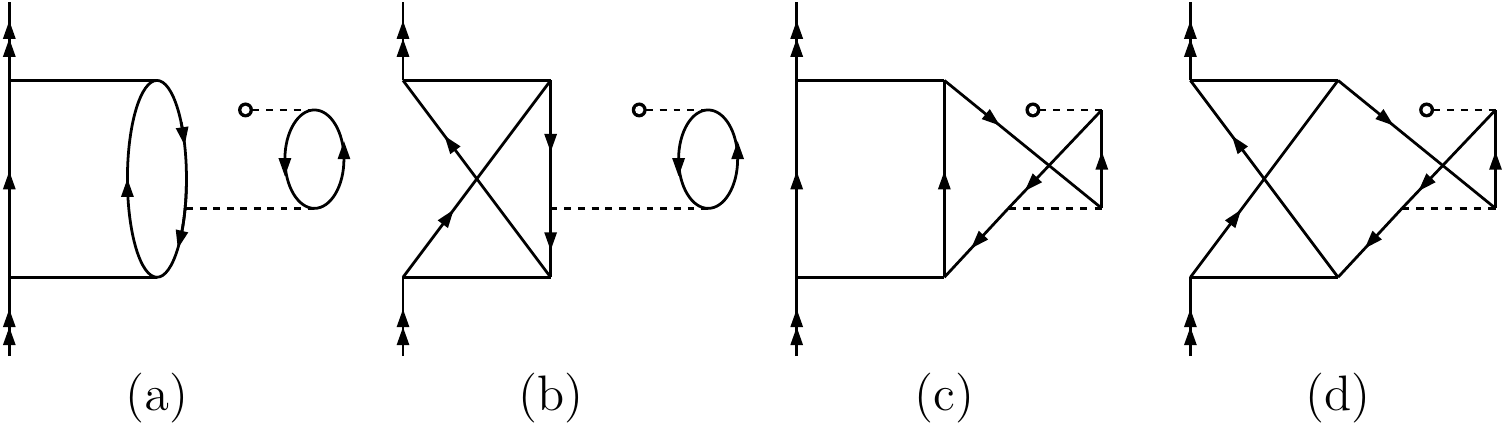}
  \caption{Goldstone HFS diagrams from 
           ${S_2}^\dagger h_{\rm hfs}S_{3({\rm cs})}$, which contribute to
           $A_{({\rm cs})}^2 $. In the first diagram (a) all the two body
           vertices, $S_2^\dagger$, $1/r_{12}$ and $S_2$ are of direct type.
           Remaining diagrams are combinations of exchange at different 
           vertices. }
  \label{hfs_cs2}
\end{center}
\end{figure}

  The triples of the $S_{3({\rm cs})}$ type have two virtual lines
above the $S_2$ vertex and no core line. This limits the number of 
allowed contractions between $h_{\rm hfs}$ and $S_2$. So, there are only two
unique topologies of the contraction
$\contraction[0.5ex]{}{h}{_{\rm hfs}}{S}h_{\rm hfs}S_{3({\rm cs})}$.
First, the core and virtual orbitals of $h_{\rm hfs} $ contract wit the
residual Coulomb interaction and $S_2$, respectively. Diagrams arising from
the contractions are shown in Fig. \ref{hfs_cs1} and contribution in
antisymmetrised notation is
\begin{equation}
   A_{({\rm cs})}^1  = \sum_{vab}\sum_{pqrs} \frac{-1}{\epsilon_{vab}^{pqr}}
       \tilde{s}_{vb}^{pr*}h_q^a\tilde{v}_{ab}^{cr}\tilde{s}_{vc}^{pq} . 
\end{equation}
And second, the $h_{\rm hfs}$ contracts with the orbital lines of residual
Coulomb interactions. There are four diagrams and are shown in 
Fig. \ref{hfs_cs2}. The contribution is
\begin{equation}
   A_{({\rm cs})}^2  = \sum_{vab}\sum_{pqrs} \frac{-1}{\epsilon_{vab}^{pqr}}
       s_{va}^{pq*}h_r^b\tilde{v}_{ab}^{cr}\tilde{s}_{vc}^{pq} . 
\end{equation}
Note, the exchange at the $S_2^\dagger$, like in $A_{({\rm vs})}^1$ and
$A_{({\rm vs})}^2$, does not generate topologically unique diagrams. 
%
%


\subsection{Contribution from $T_2^\dagger H_{\rm hfs}S_3$ }

\begin{figure}[h]
\begin{center}
  \includegraphics[width = 8.0cm]{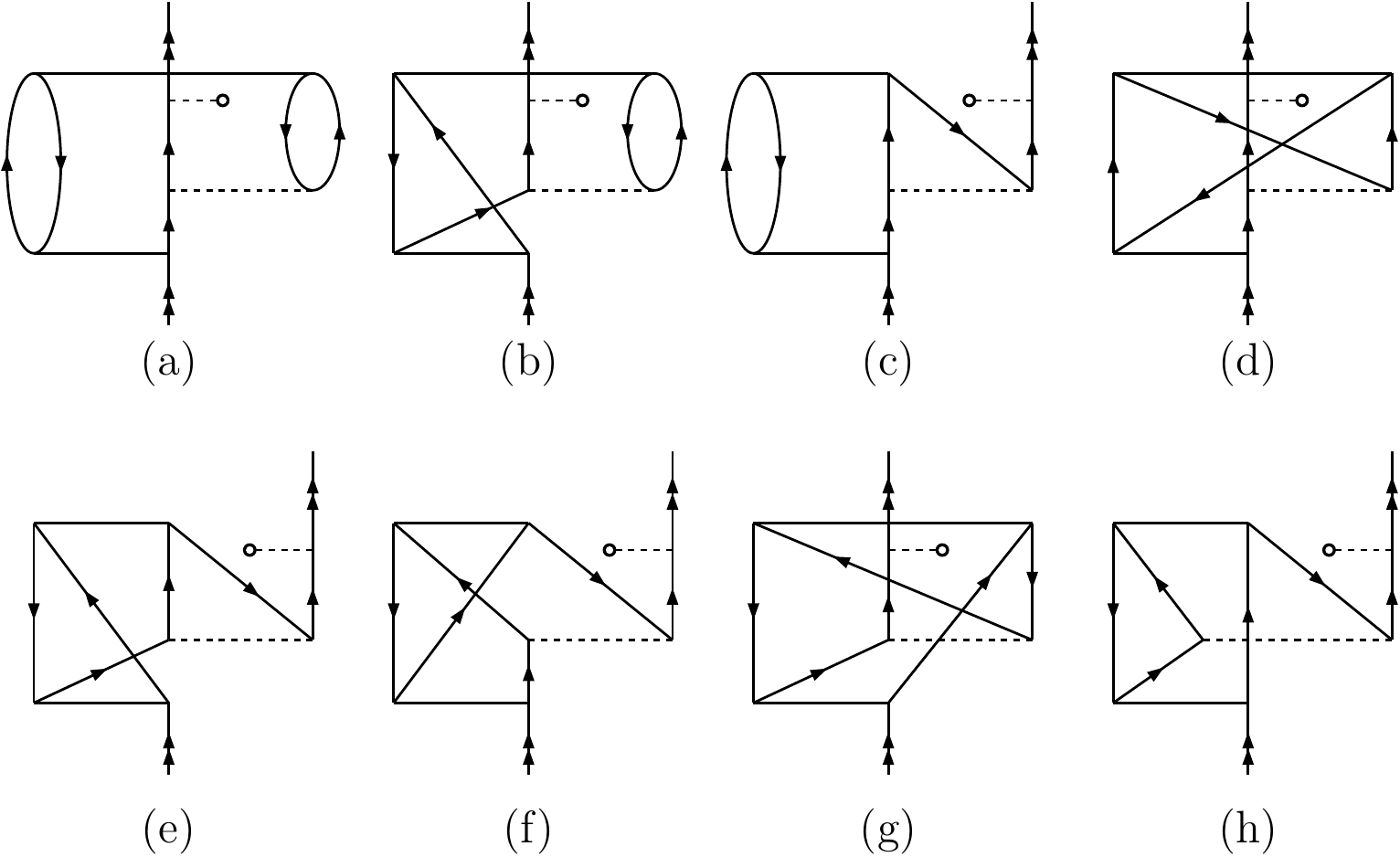}
  \caption{Goldstone HFS diagrams from 
           ${S_2}^\dagger h_{\rm hfs}S_{3({\rm vs})}$, which contribute to
           $A_{({\rm vs})}^5 $. In the first diagram (a) all the two body
           vertices, $S_2^\dagger$, $1/r_{12}$ and $S_2$ are of direct type.
           Remaining diagrams are combinations of exchange at different 
           vertices. }
  \label{hfs_vs5}
\end{center}
\end{figure}

\begin{figure}[h]
\begin{center}
  \includegraphics[width = 8.0cm]{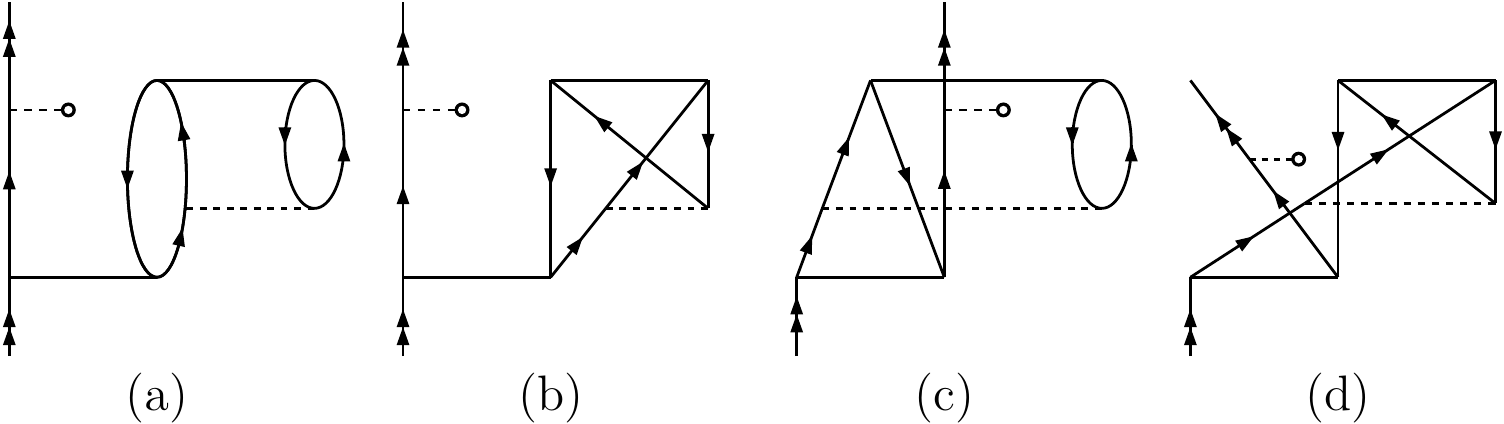}
  \caption{Goldstone HFS diagrams from 
           ${S_2}^\dagger h_{\rm hfs}S_{3({\rm vs})}$, which contribute to
           $A_{({\rm vs})}^6 $. In the first diagram (a) all the two body
           vertices, $S_2^\dagger$, $1/r_{12}$ and $S_2$ are of direct type.
           Remaining diagrams are combinations of exchange at different 
           vertices. }
  \label{a_vs6}
\end{center}
\end{figure}

 There is a key topological difference between the $T_2^\dagger H_{\rm hfs}S_3$
and $S_2^\dagger H_{\rm hfs}S_3$ diagrams. This arises from the number of
lines above vertex of the cluster operators $S_2$ and $T_2$. The former has
three, where as the later has four and more operators to contract. 
Consequently, fewer diagrams arise from $T_2^\dagger H_{\rm hfs}S_3$ and 
these, like earlier, are identified based on the topology of contractions.
Contributions from this term, like in $S_2^\dagger H_{\rm hfs}S_3$, is 
separable into $T_2^\dagger H_{\rm hfs}S_{3({\rm vs})}$ and 
$T_2^\dagger H_{\rm hfs}S_{3({\rm cs})}$. Consider the first term, there are 
two groups of diagrams. In the first group,  ${T_2}^\dagger$ contracts with 
the a pair of core and virtual lines with $S_2$ and $v$. Eight distinct 
diagrams, shown in Fig. \ref{hfs_vs5}, arise from this contraction and 
contribution is
\begin{equation}
   A_{({\rm vs})}^5  = \sum_{vab}\sum_{pqrs} \frac{1}{\epsilon_{vab}^{pqr}}
      \tilde{t}_{ab}^{qr*}h_p^v \tilde{v}_{sb}^{pr}\tilde{s}_{va}^{sq}.
\end{equation}
Here, we have given the antisymmetrised expression. The individual terms may
be written in explicit forms line in Eq. (\ref{a_vs3}). 
The second group of 
diagrams arise from the contraction of ${T_2}^\dagger$ with two virtual and 
one core orbital lines of $v$, and one core orbital line of $S_2$. Four 
diagrams arise from this term and these are given in Fig. \ref{a_vs6}. The
contribution in antisymmetrised form is
\begin{equation}
   A_{({\rm vs})}^6  = \sum_{vab}\sum_{pqrs} \frac{1}{\epsilon_{vab}^{pqr}}
      \tilde{t}_{ab}^{sr*}h_p^v v_{qb}^{sr}\tilde{s}_{va}^{pq}.
\end{equation}
Note that ${T_2}^\dagger $  and $S_2 $ are antisymmetrised in the above 
equation. Where as, $v$ and $S_2$ are antisymmetrised in the previous groups 
consisting of four diagrams. These two antisymmetrizations are equivalent and
give the same set of diagrams. The term $A_{({\rm vs})}^6$ completes the 
possible forms of HFS diagrams arising from the $S_{3({\rm vs})}$ type of 
valence triples.  Collecting all the terms, the net contribution is 
\begin{equation}
    A_{({\rm vs})} = \sum_{i=1}^6 A_{({\rm vs})}^i.
\end{equation}
To summarize, $A_{({\rm vs})} $ constitute 36 many-body Goldstone
diagrams grouped into six groups. Each group is defined based on the 
contraction topology and form of $h_{\rm hfs}$. 

\begin{figure}[h]
\begin{center}
  \includegraphics[width = 8.0cm]{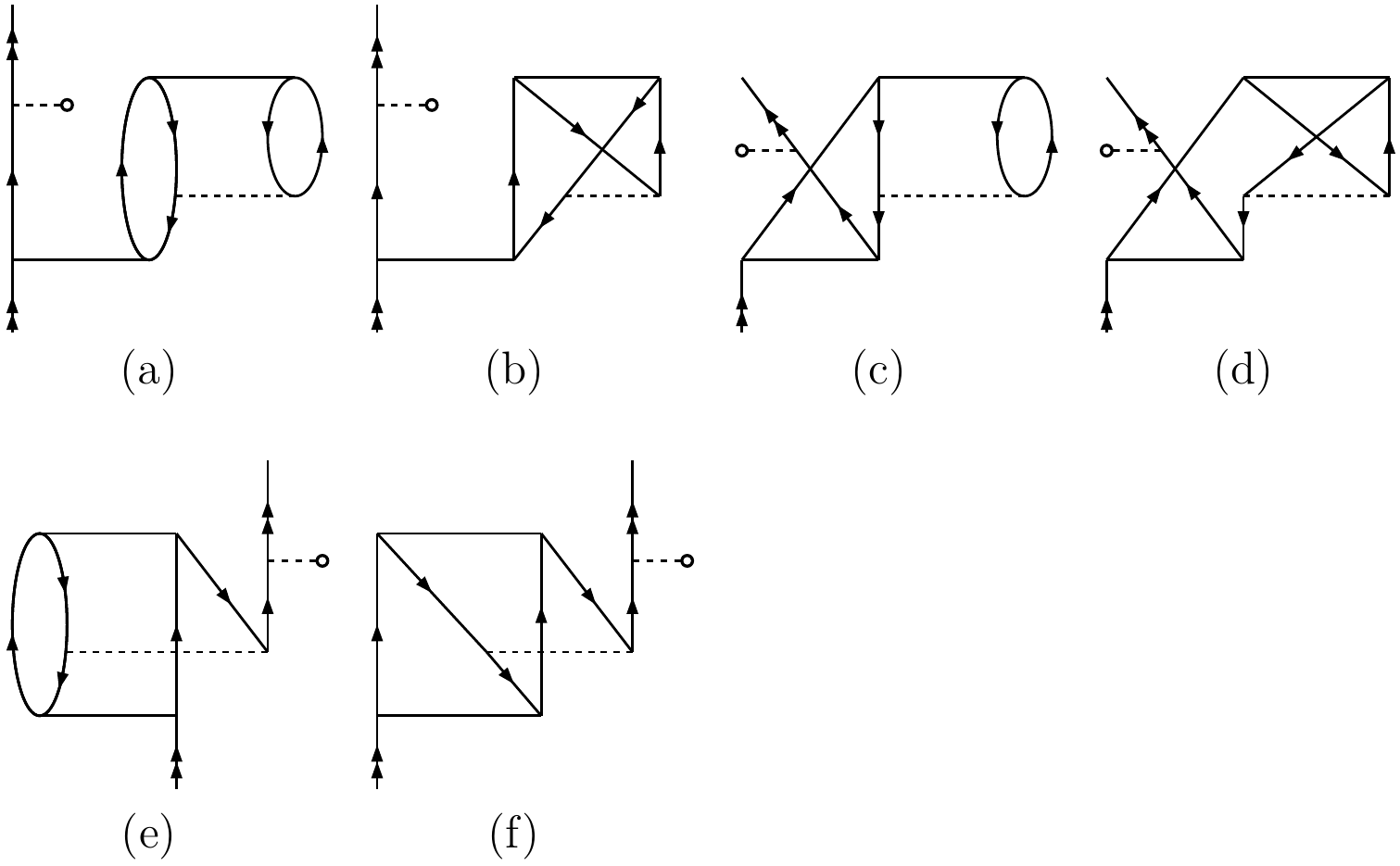}
  \caption{Goldstone HFS diagrams from 
           ${S_2}^\dagger h_{\rm hfs}S_{3({\rm cs})}$, which contribute to
           $A_{({\rm cs})}^3 $. In the first diagram (a) all the two body
           vertices, $S_2^\dagger$, $1/r_{12}$ and $S_2$ are of direct type.
           Remaining diagrams are combinations of exchange at different 
           vertices. }
  \label{a_cs34}
\end{center}
\end{figure}

From $T_2^\dagger H_{\rm hfs}S_{3({\rm cs})}$ there are two groups of diagrams.
The first group has four diagrams and these are shown in Fig. \ref{a_cs34}a-d. 
The contribution is 
\begin{equation}
   A_{({\rm cs})}^3  = \sum_{vabc}\sum_{pqr} \frac{1}{\epsilon_{vab}^{pqr}}
      \tilde{t}_{ab}^{qr*}h_p^v v_{ab}^{cr}\tilde{s}_{vc}^{pq}.
\end{equation}
The second group has two diagrams and these are shown in Fig. \ref{a_cs34}e-f.
The contribution is
\begin{equation}
   A_{({\rm cs})}^4  = \sum_{vabc}\sum_{pqr} \frac{-1}{\epsilon_{vab}^{pqr}}
      t_{ab}^{qp*}h_r^v v_{ab}^{dr}\tilde{s}_{dv}^{pq}.
\end{equation}
Collecting all the groups, the net contribution from the $S_{3({\rm cs})}$ 
type of triples is
\begin{equation}
  A_{({\rm cs})} = \sum_{i=1}^4 A_{({\rm cs})}^i.
\end{equation}
Totally there are 18 Goldstone diagrams in $A_{({\rm cs})}$. Collecting all the
diagrams from the $S_2 $ perturbed triples, we define
\begin{equation}
   A_{({\rm s})} = A_{({\rm vs})} + A_{({\rm cs})}.
\end{equation}
There are in 54 many-body diagrams and these are separable into ten groups.
This completes, excluding the structural radiation diagrams, the diagrammatic
analysis of the $S_2$ perturbed triple correction to the magnetic hyperfine 
constant.

%
%


\section{HFS constants from $T_2 $ perturbed triples}

\begin{figure}[h]
\begin{center}
  \includegraphics[width = 8.0cm]{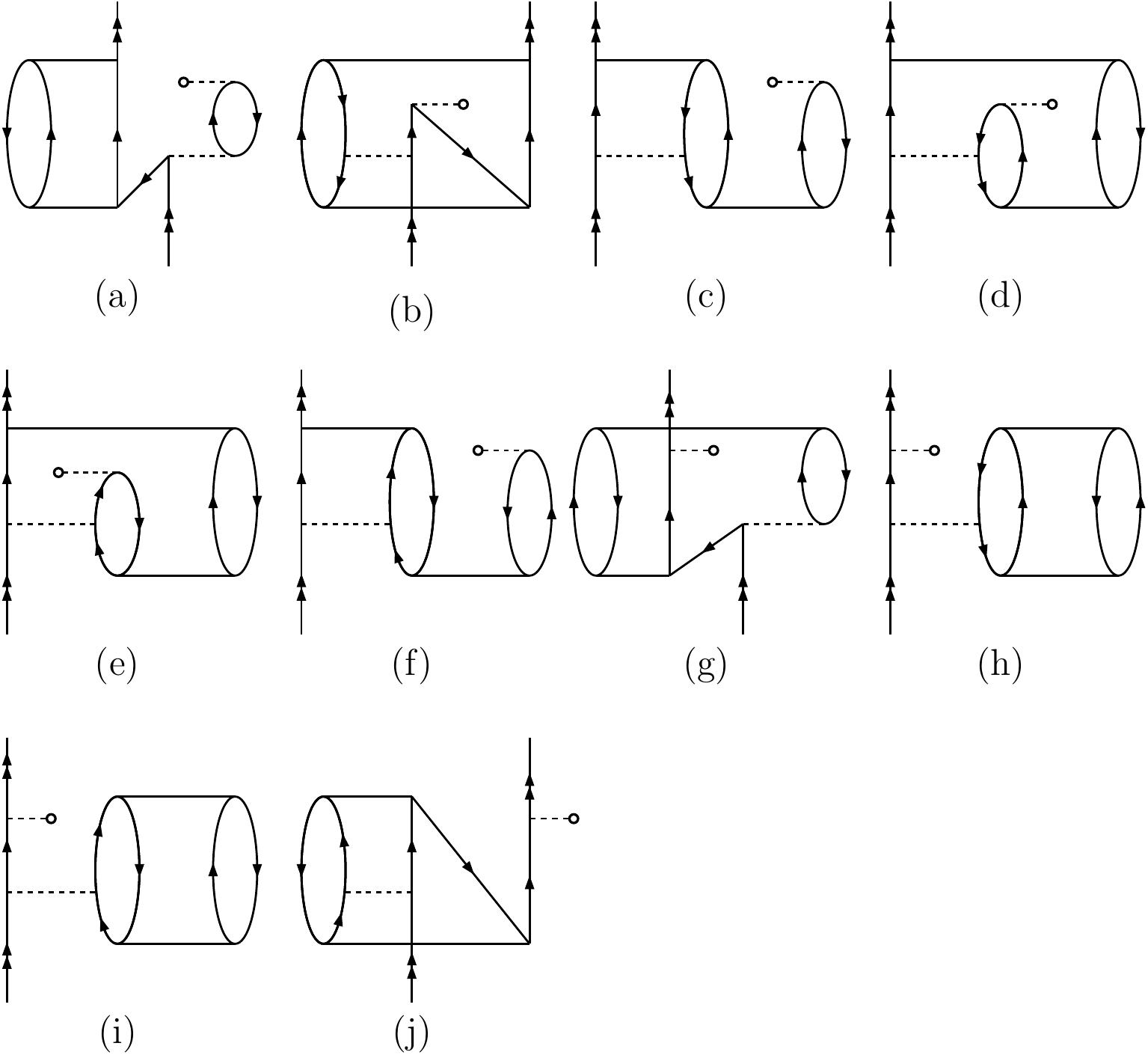}
  \caption{HFS diagrams from the $T_2$ perturbed $S_3$. Here, only the diagrams
           with direct interaction at all the interaction vertices are given.
            }
  \label{a_t2}
\end{center}
\end{figure}

  The contributions from the $T_2$ perturbed triples, like in $S_2$ perturbed
triples, is separated into two categories: perturbation to the core orbital and
virtual orbital. Contribution from these are defined as $A_{({\rm ct})}$ and
$A_{({\rm vt})}$. Similar to $A_{(\rm s)}$, diagrams from each of these
are classified into groups. Diagrams from each of the groups with direct at all
the two body vertices are given in Fig. \ref{a_t2}. The diagrams of the 
$A_{({\rm ct})}$  are shown in Fig. \ref{a_t2}(a-d) and (g-h). The expression
is
\begin{eqnarray}
   A_{({\rm ct})} & = &\sum_{vabc}\sum_{pqr} \frac{-1}{\epsilon_{vab}^{pqr}}
          \Big ( 
             s_{av}^{qp*} h_r^b\tilde{v}_{vb}^{cr}\tilde{t}_{ac}^{qp} 
           - s_{av}^{qr*} h_p^b\tilde{v}_{av}^{cp}\tilde{t}_{cb}^{qr} 
                   \nonumber \\
     &&    + \tilde{s}_{vb}^{pq*} h_r^b\tilde{v}_{vb}^{pc}\tilde{t}_{cb}^{qr} 
           + \tilde{s}_{vb}^{pr*} h_q^a\tilde{v}_{va}^{pc}\tilde{t}_{cb}^{qr} 
           + \tilde{t}_{ab}^{qr*} h_p^v\tilde{v}_{vb}^{cr}\tilde{t}_{ac}^{qp} 
                   \nonumber \\
     &&    + \tilde{t}_{ab}^{qr*} h_p^v\tilde{v}_{va}^{pc}t_{cb}^{qr} \Big ).
\end{eqnarray} 
One immediate observation is, the structure and number of the terms in the 
above expression are similar to $A_{({\rm vs})}$. Key transformations are
interchange of conversion of $S_2$ and $v$ operators to $v$ and $T_2$, 
respectively. The expression of the second category is
\begin{eqnarray}
   A_{({\rm vt})} & = &\sum_{vab}\sum_{pqrs} \frac{1}{\epsilon_{vab}^{pqr}}
          \Big ( 
             \tilde{s}_{vb}^{pr*} h_q^a\tilde{v}_{vs}^{pq}\tilde{t}_{ab}^{sr} 
           + s_{va}^{pq*} h_r^b\tilde{v}_{vs}^{pq}\tilde{t}_{ab}^{sr} 
                   \nonumber \\
     &&    + \tilde{t}_{ab}^{qr*} h_p^v\tilde{v}_{vs}^{pq}\tilde{t}_{ab}^{sr} 
           - t_{ab}^{qp*} h_r^v\tilde{v}_{sv}^{qp}t_{ab}^{sr} \Big ).
\end{eqnarray}
Here, the terms are similar to $A_{({\rm cs})} $ and same transformations 
discussed in $A_{({\rm ct})}$ apply.
 
\begin{table}[h]
\caption{Basis set parameters $\alpha$ and $\beta$ used in the 
         calculations.}
  \label{tab:basis-param}
\begin{ruledtabular}
\begin{tabular}{ccccc}
Atom & Orbital & $\alpha$ & $\beta$ & Basis function \\
 \hline
Rb     & $s$ & $0.00521$ & $2.9500$ & $33$ \\
       & $p$ & $0.00655$ & $2.9950$ & $30$ \\
       & $d$ & $0.00654$ & $2.9720$ & $28$ \\
       &&&& \\
Sr$^+$ & $s$ & $0.00825$ & $2.9000$ & $35$ \\
       & $p$ & $0.00715$ & $2.9450$ & $32$ \\
       & $d$ & $0.00730$ & $2.9100$ & $30$ \\
\end{tabular}
\end{ruledtabular}
\end{table}


\section{Results and discussions}

\subsection{Single particle states}

The first step of our calculations, like in any atomic many-body calculations,
is to solve the single particle eigenvalue equations with Dirac-Hartree-Fock 
potential. For this, we consider the nuclear potential $V_N(\mathbf{r})$ 
arising from the finite size Fermi density distribution
\begin{equation}
  \rho_{\rm nuc}(r) = \frac{\rho_0}{1 + e^{(r-c)/a} },
\end{equation}
here, $a = t 4\ln 3$. The parameter $c$ is the half-charge radius, that is
$\rho_{\rm nuc}(c)=\rho_0/2$ and $t$ is the skin thickness. At the single
particle level, the spin orbitals are of the form
\begin{equation}
  \psi_{n\kappa m}(\mathbf{r})=\frac{1}{r}
  \left(\begin{array}{r}
            P_{n\kappa}(r)\chi_{\kappa m}(\hat{\mathbf{r}})\\
           iQ_{n\kappa}(r)\chi_{-\kappa m}(\hat{\mathbf{r}})
       \end{array}\right),
  \label{spin-orbital}
\end{equation}
where $P_{n\kappa}(r)$ and $Q_{n\kappa}(r)$ are the large and small component
radial wave functions, $\kappa$ is the relativistic total angular momentum
quantum number and $\chi_{\kappa m}(\hat{\mathbf{r}})$ are the spin or spherical
harmonics. One representation of the radial components is to define these
as linear combination of Gaussian like functions and are referred to as
Gaussian type orbitals (GTOs). Then, the large and small
components \cite{mohanty-89,chaudhuri-99} are
\begin{eqnarray}
   P_{n\kappa}(r) = \sum_p C^L_{\kappa p} g^L_{\kappa p}(r),  \nonumber \\
   Q_{n\kappa}(r) = \sum_p C^S_{\kappa p} g^S_{\kappa p}(r).
\end{eqnarray}
The index $p$ varies over the number of the basis functions.
For large component we choose
\begin{equation}
  g^L_{\kappa p}(r) = C^L_{\kappa i} r^{n_\kappa} e^{-\alpha_p r^2},
\end{equation}
here $n_\kappa$ is an integer. Similarly, the small component are
derived from the large components using kinetic balance condition. The
exponents in the above expression follow the general relation
\begin{equation}
  \alpha_p = \alpha_0 \beta^{p-1}.
  \label{param_gto}
\end{equation}
The parameters $\alpha_0$ and $\beta$ are optimized for each of the ions to
provide good description of the properties. In our case the optimization is to
reproduce the numerical result of the total and orbital energies. The
optimized parameters used in the calculations are listed in
Table.\ref{tab:basis-param}.
\begin{table}[h]                
\caption{Excitation energies calculated using RCC, compared with
         other theoretical results and experimental data.      
         All values are in atomic units.}                            
\begin{ruledtabular}                                                 
\begin{tabular}{ccccc}                                               
Atom & State & This work & Other works & Exp {Ref\cite{nist}.}\\
\hline                                                                  
$^{85}$Rb                                                               
      &$5s_{1/2}$ & $0.0$     & $0.0$ & $0.0$                   \\
      &$5p_{1/2}$ & $0.05759$ & $0.05718^{\rm a}$ & $0.05731$   \\
      &$5p_{3/2}$ & $0.05872$ & $0.05826^{\rm a}$ & $0.05840$   \\
      &$4d_{3/2}$ & $0.08836$ & $0.08822^{\rm a}$ & $0.08819$   \\
      &$4d_{5/2}$ & $0.08836$ & $0.08820^{\rm a}$ & $0.08819$   \\
                                               \\                
$^{87}$Sr$^+$                                                   
      &$5s_{1/2}$ & $0.0$     & $0.0$ & $0.0$                   \\
      &$5p_{1/2}$ & $0.10841$ & $0.11001^{\rm b}$ & $0.10805$   \\
      &$5p_{3/2}$ & $0.11221$ & $0.11376^{\rm b}$ & $0.11171$   \\
      &$4d_{3/2}$ & $0.06611$ & $0.06560^{\rm b}$ & $0.06632$   \\      
      &$4d_{5/2}$ & $0.06724$ & $0.06707^{\rm b}$ & $0.06760$   \\
\end{tabular}
\end{ruledtabular}
\begin{flushleft}
$^{\rm a}$ Reference\cite{safronova-11}
$^{\rm b}$ Reference\cite{guet-91}
\end{flushleft}                                                                 
\end{table}                                                                  

  For Rb and Sr$^+$ we use $V^{N-1}$  and $V^{N-2}$ orbitals, respectively.
These are the single particle eigenfunctions of the Rb$^+$ and Sr$^{2+}$ ions,
respectively. The single particle basis sets have few bound
states and rest are continuum. We optimize the basis such that: single
particle energies of the core and valence orbitals are in good agreement with
the numerical results. For this we use GRASP92 \cite{parpia-96} to generate
the numerical results. It is to be noted that, the basis parameters in 
Table.\ref{tab:basis-param} are different from one give in our earlier work
\cite{mani-10}. Between the two, the present is better optimized and of higher
quality. The optimization is nontrivial as there are several parameters and
single particle equations are solved self-consistently.

   From Eq.(\ref{spin-orbital}) the reduced matrix element of the magnetic
hyperfine operator between two spin orbitals , $v'$ and $v$, is
\begin{eqnarray}
  \langle v'\red  t^1\red v\rangle &=& -(\kappa_v + \kappa_{v'})
  \langle -\kappa_{v'}\red C^1\red\kappa_v \rangle \nonumber \\
  &&\times \int^\infty_0 \frac{dr}{r^2}(P_{n_{v'}\kappa_{v'}} Q_{n_v\kappa_v}
       + Q_{n_{v'}\kappa_{v'}} P_{n_v\kappa_v}). \;\;\;\;\;\;
  \label{hfs_matrix}
\end{eqnarray}
A detailed derivation is given in Ref. \cite{johnson-07}.


\subsection{Cluster amplitudes and normalization}

 As described in our earlier works \cite{mani-09,mani-10,mani-11}, the RCC 
equations are solved iteratively using the Jacobi method. To improve 
convergence we employ direct inversion in the iterated subspace (DIIS) 
\cite{pulay-80}. The cluster amplitudes are solved for each Hilbert space
manifold of the total Fock space. At each step the Hilbert space is augmented
with one electron. In short, the $T$ equations are solved first and these are 
used to generate the open shell cluster amplitudes $S$. 

 One important point is, the cluster equations are in terms of the reduced
matrix elements. So the solutions are independent of magnetic quantum  numbers
and appropriate phase factors are required to define the cluster amplitudes
in the cojugate manifold and these are
\begin{eqnarray}
    t_a^{p*} &=& (-1)^{j_p-j_a}t_a^p, \\
    t_{ab}^{pq*} &=& (-1)^{j_p + j_q -j_b-j_a}t_{ab}^{pq}.
\end{eqnarray}
These relations apply in any calculation which involve $T^\dagger$ and
$S^\dagger$. The coupled-cluster wave function is normalized and the 
normalization factor is
\begin{equation}
     {\cal N} = \langle \Psi_v|\Psi_v\rangle = 
     \langle \Phi_v|e^{T^\dagger}\left (1 + S^\dagger \right)
     \left (1 + S \right ) e^{T}|\Phi_v\rangle .
\end{equation}
Here, $e^{T^\dagger}e^{T}$ is a non-terminating operator. For the present we
consider the approximation
\begin{equation}
  {\cal N} \approx \langle \Phi_v|1 + S^\dagger S + T^\dagger T|\Phi_v\rangle .
\end{equation}
The higher order terms 
${(T_1^\dagger)}^\alpha {(T_2^\dagger)}^\beta{(T_1)}^\gamma {(T_2)}^\delta$,
such that
$\alpha + \beta > 1$, $\gamma + \delta  > 1$ and  
$\alpha + \beta - \gamma - \delta = 0$, are neglected. We also neglect the 
mixed operator term $(T_1^\dagger S_2  + \textrm{c.c.})$  and higher orders.


\subsection{Excitation energies}
 To determine the quality of the basis set and parameters, we compute the
attachment energies of the ground state ($S_{1/2}$) and the first
excited $P_{1/2}$, $P_{3/2}$, $D_{3/2}$ and $D_{5/2}$ states are calculated.
Then the ionization potential (IP), the energy required to remove the valence
electron, is the negative of the attachement energy $ -E^{\rm att}$.
To calculate the excitation energy (EE) of the state $|\Psi_v\rangle $,
consider $E^{\rm att}_g$ and $ E^{\rm att}_v$ as the attachment
energies of the ground state and excited state. Then difference
$ E^{\rm att}_v- E^{\rm att}_g$ is the EE, it can as well be
defined in terms of IPs.
\begin{table*}               
\caption{Magnetic dipole hyperfine structure constants for
         $^{85}$Rb and $^{87}$Sr$^+$. The values given are in the unit 
         of MHz.}                                                      
\label{tab-hfs}                                                        
\begin{ruledtabular}                                                   
\begin{tabular}{cccccc}                                                
Atom & State & \multicolumn{2}{c}{This work} & Other works & Experiment \\
\hline                                                                    
     &     &   CCSD    &  CCSD(T)    &             &                    \\
\hline
$^{85}$Rb
&$5s_{1/2}$ & $1030.94$ & $1030.60$   & $1011.1^{\rm a}$
                                      & $1011.91(2)^{\rm b}$ \\
&$5p_{1/2}$ & $120.69$  & $120.67$    & $120.4^{\rm a}$
                                      & $120.72(25)^{\rm b}$ \\
&$5p_{3/2}$ & $24.48$   & $24.63$     & $24.5^{\rm a}$ 
                                      & $24.99(1)^{\rm c}$ \\
&$4d_{3/2}$ & $7.85$    & $7.80$      & $-$
                                      & $7.3(35)^{\rm d},7.329(35)^{\rm e}$ \\
&$4d_{5/2}$ & $-4.78$   & $-4.77$     & $-$
                                      & $-5.2(3)^{\rm d}$ \\
$^{87}$Sr$^+$
&$5s_{1/2}$ & $-1014.20$ & $-1013.86$ & $-10003.18^{\rm f},-1000^{\rm g}$
                                      & $1000.5(1.0)^{\rm j}$ \\
&$5p_{1/2}$ & $-178.73$  & $-178.67$  & $-178.40^{\rm f},-177^{\rm g},
                                         -175^{\rm h}$ & $-$ \\
&$5p_{3/2}$ & $-35.28$   & $-35.48$   & $-35.11^{\rm f},-35.3^{\rm g},
                                         -30^{\rm h}$ & $-36.0^{\rm j}$ \\
&$4d_{3/2}$ & $-46.30$   & $-46.21$   & $-47.36^{\rm f},-46.7^{\rm g}$
                                      & $-$ \\
&$4d_{5/2}$ & $1.71$     & $1.71$     & $2.51^{\rm f},1.1^{\rm g},2.156^{\rm i}$
                                      & $2.17^{\rm k}$ \\
\end{tabular}                                                                 
\end{ruledtabular}                                                            
\begin{flushleft}                                                               
$^{\rm a}$ Reference\cite{safronova-99},
$^{\rm b}$ Reference\cite{arimondo-77},
$^{\rm c}$ Reference\cite{arimondo-75},
$^{\rm d}$ Reference\cite{lam-80},
$^{\rm e}$ Reference\cite{moon-09},
$^{\rm f}$ Reference\cite{yu-04},
$^{\rm g}$ Reference\cite{martensson-02}, \\
$^{\rm h}$ Reference\cite{heully-85}, 
$^{\rm i}$ Reference\cite{sahoo-07},
$^{\rm j}$ Reference\cite{buchinger-90},
$^{\rm k}$ Reference\cite{barwood-03}.                                        
\end{flushleft}                                                               
\end{table*}  

\begin{table*}[ht]                        
\caption{The HFS contributions from different terms in RCC. 
         All the values are in MHz.}                             
\label{tab-hfs-comp}                                                
\begin{ruledtabular}                                                
\begin{tabular}{cccccccccc}                                          
Atom& State &\multicolumn{7}{c}{RCC terms}                               \\ 
\hline                                                                   \\ 
   &       & DF & $\tilde H_{\rm hfs}$-DF & $S^\dagger\tilde H_{\rm hfs}$
           & $S^\dagger_2\tilde H_{\rm hfs} S_1$
           & $S^\dagger_1\tilde H_{\rm hfs} S_1$
           & $S^\dagger_2\tilde H_{\rm hfs} S_2$ & Other & Norm            \\
   &       &    &               & $+ c.c$ & $+ c.c.$ & & & &               \\
 \hline                                                                    \\
$^{85}$Rb  &$5s_{1/2}$&$646.003$&$-12.115$&$375.719$&$19.822$&$23.564$&     
                                           $20.498$&$-22.231$&$1.020$      \\
      &$5p_{1/2}$&$69.883$&$-0.762$&$47.411$ &$2.762$&$2.751$&$1.630$&   
                                              $-1.760$&$1.010$             \\
      &$5p_{3/2}$&$12.417$&$-0.054$&$10.919$ &$0.804$&$0.450$&$0.479$&   
                                              $-0.312$&$1.009$             \\
      &$4d_{3/2}$&$3.116$&$0.127$&$3.153$ &$0.119$&$1.014$&$0.685$&   
                                              $-0.068$&$1.039$             \\
      &$4d_{5/2}$&$1.328$&$0.087$&$-4.694$ &$-2.550$&$0.420$&$0.472$&
                                              $-.028$&$1.038$              \\
                      \\
$^{87}$Sr$^+$&$5s_{1/2}$&$-735.629$&$8.044$&$-284.856$&$-9.987$&$-8.594$&     
                                          $-15.624$&$16.907$&$1.015$       \\
      &$5p_{1/2}$&$-122.123$&$1.369$&$-56.502$ &$-2.155$&$-1.862$&$-1.566$&   
                                                       $2.217$&$1.010$     \\
      &$5p_{3/2}$&$-21.449$&$0.191$&$-13.023$ &$-0.582$&$-0.307$&$-0.869$&   
                                                        $0.412$&$1.010$    \\
      &$4d_{3/2}$&$-31.401$&$-0.803$&$-12.400$ &$-0.594$&$-0.428$&$-2.020$&   
                                                         $0.283$&$1.023$   \\
      &$4d_{5/2}$&$-13.091$&$-0.428$&$16.306$ &$1.051$&$-0.177$&$-2.018$&   
                                                     $0.116$&$1.023$       \\
\end{tabular}                                                                  
\end{ruledtabular}                                                             
\end{table*}

\subsection{HFS constants}

 To compute the hyperfine constants from the CCSD wave functions, we use
Eq.(\ref{hfs_cc}). The results are listed in Table.\ref{tab-hfs}, for
comparison the results of other theoretical calculations and experimental
data are also given. As defined in Eq.(\ref{hfs_cc}), the coupled-cluster
expression of the hyperfine structure constants is separated into three
groups. The dominant contribution from the first term $\tilde H_{\rm hfs}$,
up to first order in $T^\dagger$ and $T$, is
\begin{equation}
  \tilde H_{\rm hfs} \approx H_{\rm hfs} + 2H_{\rm hfs}T_1 + 
     T_1^\dagger H_{\rm hfs}\left ( T_1 + 2T_2 \right ) + 
     T_2^\dagger H_{\rm hfs}T_2 .
\end{equation}
Here,  the first term is the Dirac-Fock (DF), which has the largest
contribution. The factor two in the second and fourth terms accounts for the
complex conjugate terms. The third term, second order in $T_1$, has one
diagram and negligibly small contribution. The diagrams arising from the last
term are topologically are the structural radiation diagrams and have
neglible contributions. Topologically, these are insertion of $H_{\rm hfs}$ to 
the normalization diagrams and contribution from these are labelled as
$\tilde H_{\rm hfs}-{\rm DF}$. Detailed diagrammatic analysis are given in 
our previous work \cite{mani-10}. The last two terms in Eq.(\ref{hfs_cc}) are
approximated as
\begin{eqnarray}
  S^\dagger \tilde H_{\rm hfs}   & \approx &  2S^\dagger  
              \left ( H_{\rm hfs}e^T\right )_1 ,  \\ 
  S^\dagger \tilde H_{\rm hfs} S & \approx &    S_1^\dagger H_{\rm hfs}\left 
              ( S_1 + 2S_2 \right ) + 
     S_2^\dagger H_{\rm hfs}S_2 .
\end{eqnarray}
Like in $\tilde H_{\rm hfs} $, the factor of two is to account for the complex
conjugate terms. Based on this grouping, the contributions are listed in 
Table.\ref{tab-hfs-comp}. In the following we present a detailed comparison 
of  our magnetic hyperfine constants results with the earlier ones. As 
discussed later, some of our results are the best match with experimental 
data.


\subsection{HFS constants contribution from triples}

  The HFS constants after including the perturbed triples are listed in the 
Table. \ref{tab-hfs}. There is negligible contribution for $S_{1/2}$ and 
$P_{1/2}$ states. These are 0.03\% and 0.02\% for Rb and, 0.03\% and 0.03\% 
for Sr$^+$. However, for $P_{3/2}$, $D_{3/2}$, and $D_{5/2}$ contributions 
from triples are not small and could be important in high precision atomic 
theory calculation. And these are 0.6\%, 0.6\% and 0.2\% for Rb, and 0.6\%, 
0.2\% and 0.5\% for Sr$^+$. 
The observed pattern of perturbed triples contribution is different from
$^{87}$Rb reported in Ref. \cite{safronova-11}. This could be
on account of two factors: difference in the nature of the single particle
basis functions, and isotope specific effects. The later may not be 
the dominant cause as the electron wavefunctions have little variation
for isotopes of small mass differences.                  
For Sr$^+$ on the other hand,  there are no previous theoretical work on the
effects of triple excitations. However, our results exhibits trends similar 
to previous work on  Ca$^+$ \cite{sahoo-09}, which reported
the contributions from triples as 0.002\%, 0.08\%, 0.10\%,           
0.11\% and 0.29\% for $S_{1/2}$, $P_{1/2}$, $P_{3/2}$, $D_{3/2}$, and          
$D_{5/2}$ states, respectively.                                                
%
%
\begin{table}[ht]
\caption{The term wise contribution of HFS constants from triples for Rb.
         All values listed are in atomic units.} 
\label{tab-hfs-rb}
\begin{ruledtabular}
\begin{tabular}{crrrrr}
RCC term & $5s_{1/2}$ & $5p_{1/2}$ & $5p_{3/2}$ & $4d_{3/2}$ & $4d_{5/2}$ \\
\hline 
$A_{({\rm vs})}^1$ & $0.083$  & $0.013$  & $0.013$  & $-0.003$ & $-0.054$ \\
$A_{({\rm vs})}^2$ & $0.027$  & $0.000$  & $0.014$  & $0.004$  & $0.002$  \\
$A_{({\rm vs})}^3$ & $0.200$  & $0.003$  & $0.024$  & $0.020$  & $0.000$  \\
$A_{({\rm vs})}^4$ & $0.064$  & $-0.002$ & $0.012$  & $0.000$  & $-0.028$ \\
$A_{({\rm vs})}^5$ & $-0.236$ & $-0.017$ & $-0.010$ & $-0.002$ & $-0.002$ \\
$A_{({\rm vs})}^6$ & $-0.135$ & $-0.014$ & $-0.008$ & $-0.005$ & $-0.004$ \\
     \hline
$A_{({\rm vs})}  $ & $0.003$  & $-0.017$ & $0.045$  & $0.014$  & $-0.086$ \\
     &&&&&\\
$A_{({\rm cs})}^1$ & $-0.141$ & $-0.007$ & $0.024$  & $-0.018$ & $0.058$  \\
$A_{({\rm cs})}^2$ & $0.012$  & $0.007$  & $0.000$  & $0.001$  & $0.015$  \\
$A_{({\rm cs})}^3$ & $0.216$  & $0.018$  & $0.010$  & $0.007$  & $0.005$  \\
$A_{({\rm cs})}^4$ & $-0.037$ & $-0.002$ & $-0.001$ & $-0.001$ & $-0.001$ \\
     \hline
$A_{({\rm cs})}  $ & $0.050$  & $0.016$  & $0.033$  & $-0.011$ & $0.077$  \\
     &&&&&\\
$A_{({\rm ct})}^1$ & $-0.219$ & $-0.006$ & $-0.005$ & $0.000$  & $0.000$  \\
$A_{({\rm ct})}^2$ & $-0.007$ & $-0.002$ & $0.000$  & $-0.005$ & $-0.001$ \\
$A_{({\rm ct})}^3$ & $-0.005$ & $0.000$  & $-0.002$ & $0.001$  & $-0.003$ \\
$A_{({\rm ct})}^4$ & $-0.007$ & $-0.002$ & $-0.002$ & $-0.002$ & $-0.007$ \\
$A_{({\rm ct})}^5$ & $-0.149$ & $-0.015$ & $-0.007$ & $-0.011$ & $-0.007$ \\
$A_{({\rm ct})}^6$ & $1.966$  & $0.195$  & $0.113$  & $0.035$  & $0.030$  \\
     \hline
$A_{({\rm ct})}  $ & $1.579$  & $0.170$  & $0.097$  & $0.018$  & $0.012$  \\
     &&&&&\\
$A_{({\rm vt})}^1$ & $-0.084$ & $-0.006$ & $0.002$  & $-0.014$ & $0.007$  \\
$A_{({\rm vt})}^2$ & $0.008$  & $-0.001$ & $-0.002$ & $-0.003$ & $0.014$  \\
$A_{({\rm vt})}^3$ & $-1.782$ & $-0.178$ & $-0.099$ & $-0.028$ & $-0.023$ \\
$A_{({\rm vt})}^4$ & $0.054$  & $0.006$  & $0.004$  & $0.004$  & $0.003$  \\
     \hline
$A_{({\rm vt})}  $ & $-1.804$ & $-0.179$ & $-0.095$ & $-0.041$ & $0.001$  \\
\end{tabular}
\end{ruledtabular}
\end{table}


\subsubsection{Rb}

   In the Table. \ref{tab-hfs-rb}, we have listed the individual contributions 
from the different groups of triples HFS diagrams for Rb. In all the cases we 
notice large cancellations. For $S_{1/2}$ and $P_{1/2}$ states, the leading
order (LO) and next to leading order (NLO) contribution arise from
$A^6_{\rm ct}$ and $A^3_{\rm vt}$. Each of these terms contribute
$\approx$ 0.2\%.  However, the two are of opposite signs and cancel each      
other. The other dominant contributing terms are $A^5_{\rm vs}$ and
$A^3_{\rm cs}$ with contrbutions  $\approx$ 0.02\%. The two contributions 
are opposite in sign and like in LO and NLO, there are large cancellations.  

   For the state $P_{3/2}$, $A^6_{\rm ct}$ and $A^3_{\rm vt}$ are again 
the LO and NLO terms with contributions of $\approx$ 0.5\% and 
$\approx$ 0.4\%, respectively. The two contributions are of opposite sign
and nearly cancel. Other dominant contributions arise from $A^3_{\rm vs}$ and 
$A^1_{\rm cs}$, each of the contributions are $\approx$ 0.1\%. Unlike the 
cases considered and discussed so far, the two are of same phase.  For 
$D_{3/2}$ too like in $S_{1/2}$, $P_{1/2}$ and $P_{3/2}$, $A^6_{\rm ct}$ and 
$A^3_{\rm vt}$ are the LO and NLO terms. Contributions from these terms are
$\approx$ 0.4\%, but opposite in sign. The other dominant contributions are 
from $A^3_{\rm vs}$ and $A^1_{\rm cs}$ and these are $\approx$ 0.3\% and 
$\approx$ 0.2\%, respectively. Like in LO and NLO these are of opposite sign
and nearly cancel. 

   The state $D_{5/2}$ shows a different pattern of contributions. Unlike the 
states discussed so far, the dominant LO term is $A^1_{\rm cs}$ and the 
contribution is about 1.2\%. The NLO term is $A^1_{\rm vs}$ and has a 
contribution  of $\approx $1.1\%. and is opposite to the LO term. The 
terms $A^6_{\rm ct}$ and $A^3_{\rm vt}$, which are the LO and NLO of $S_{1/2}$,
$P_{1/2}$, $P_{3/2}$ and $D_{3/2}$, are the third and fourth dominant terms. 
Contributions from these terms are $\approx$ 0.6\% and $\approx$ 0.5\%, 
respectively.
%
%
\begin{table}[t]
\caption{The term wise contribution of HFS constants from triples for Sr$^+$.
         All values listed are in atomic units.} 
\label{tab-hfs-sr}
\begin{ruledtabular}
\begin{tabular}{crrrrr}
RCC term & $5s_{1/2}$ & $5p_{1/2}$ & $5p_{3/2}$ & $4d_{3/2}$ & $4d_{5/2}$ \\
\hline 
$A_{({\rm vs})}^1$ & $-0.066$ & $-0.016$ & $-0.015$ & $-0.011$ & $0.098$  \\
$A_{({\rm vs})}^2$ & $-0.020$ & $0.001$  & $-0.020$ & $-0.005$ & $0.001$  \\
$A_{({\rm vs})}^3$ & $-0.129$ & $0.001$  & $-0.039$ & $-0.046$ & $0.024$  \\
$A_{({\rm vs})}^4$ & $-0.045$ & $0.000$  & $-0.015$ & $-0.001$ & $0.070$  \\
$A_{({\rm vs})}^5$ & $0.212$  & $0.027$  & $0.016$  & $0.012$  & $0.010$  \\
$A_{({\rm vs})}^6$ & $0.109$  & $0.018$  & $0.011$  & $0.026$  & $0.020$  \\
     \hline
$A_{({\rm vs})}$   & $0.061$  & $0.031$  & $-0.062$ & $-0.025$ & $0.223$  \\
     &&&&&\\
$A_{({\rm cs})}^1$ & $0.092$  & $0.008$  & $-0.028$ & $0.036$  & $-0.098$ \\
$A_{({\rm cs})}^2$ & $-0.010$ & $-0.004$ & $0.000$  & $-0.010$ & $-0.055$ \\
$A_{({\rm cs})}^3$ & $-0.157$ & $-0.022$ & $-0.012$ & $-0.032$ & $-0.024$ \\
$A_{({\rm cs})}^4$ & $0.028$  & $0.003$  & $0.003$  & $0.006$  & $0.004$  \\
     \hline
$A_{({\rm cs})}$   & $-0.047$ & $-0.015$ & $-0.037$ & $0.000$  & $-0.173$ \\
     &&&&&\\
$A_{({\rm ct})}^1$ & $0.158$  & $0.007$  & $0.008$  & $0.000$  & $0.000$  \\
$A_{({\rm ct})}^2$ & $0.006$  & $0.004$  & $0.002$  & $0.016$  & $0.002$  \\
$A_{({\rm ct})}^3$ & $0.003$  & $0.000$  & $0.004$  & $-0.001$ & $0.006$  \\
$A_{({\rm ct})}^4$ & $0.005$  & $0.002$  & $0.001$  & $0.005$  & $0.024$  \\
$A_{({\rm ct})}^5$ & $0.107$  & $0.017$  & $0.008$  & $0.092$  & $0.059$  \\
$A_{({\rm ct})}^6$ & $-1.935$ & $-0.302$ & $-0.172$ & $-0.327$ & $-0.266$ \\
     \hline
$A_{({\rm ct})}$   & $-1.656$ & $-0.272$ & $-0.149$ & $-0.215$ & $-0.175$ \\
     &&&&&\\
$A_{({\rm vt})}^1$ & $0.069$  & $0.010$  & $-0.003$ & $0.038$  & $-0.030$ \\
$A_{({\rm vt})}^2$ & $-0.003$ & $0.001$  & $0.002$  & $0.010$  & $-0.032$ \\
$A_{({\rm vt})}^3$ & $1.786$  & $0.281$  & $0.155$  & $0.268$  & $0.209$  \\
$A_{({\rm vt})}^4$ & $-0.041$ & $-0.008$ & $-0.004$ & $-0.033$ & $-0.025$ \\
     \hline
$A_{({\rm vt})}$ &   $1.811$  & $0.284$  & $0.150$  & $0.283$  & $0.122$  \\
\end{tabular}
\end{ruledtabular}
\end{table}


\subsubsection{Sr$^+$}

   The contributions from different RCC terms for Sr$^+$ are as listed
in the Table. \ref{tab-hfs-sr}. For the states $S_{1/2}$, $P_{1/2}$ and
$P_{3/2}$, the LO and NLO have the same pattern as  Rb is observed. The only
difference is, signs are reversed. For example, in the case of $S_{1/2}$
of Rb, $A^3_{\rm cs}$ and $A^6_{\rm ct}$ are positive in sign,  and
$A^5_{\rm vs}$ and $A^3_{\rm vt}$ have negative sign. The pattern is opposite 
in the case of Sr$^+$. The reason is trivial and is on account of the opposite 
sign of $\mu$, which for Rb and Sr$^+$ are positive and negative, respectively. 

   For $D_{3/2}$ state, the LO and NLO contributions are large in comparison 
to the case of Rb. The LO and NLO terms are $A^6_{\rm ct}$ and $A^3_{\rm vt}$, 
respectively, and contributions are $\approx$ 0.7\% and $\approx$ 0.6\%.
The other dominant contributions are from $A^3_{\rm vs}$ and $A^1_{\rm cs}$, 
and these are smaller than Rb. 

  The state $D_{5/2}$ show the largest LO and NLO contributions among all the 
states. These are 15.5\% and 12.2\% from the terms $A^6_{\rm ct}$ and 
$A^3_{\rm vt}$, respectively. However, the total contribution is 3.3\% as they 
are of opposite signs. The other dominant contribution arise from 
$A^1_{\rm vs}$ and $A^1_{\rm cs}$. Contributions from each of these terms 
are 5.7\% but are opposite in sign.

%
%
\begin{table}[t]                
\caption{Excitation energies calculated using RCC for different sets
         of basis functions. All values are in atomic units.}
\label{ee_sets}
\begin{ruledtabular}                                                 
\begin{tabular}{ccccccc}                                               
Atom & State & Set 1 & Set 2  & Set 3 & Set 4 & Set 5 \\
\hline                                                                  
$^{85}$Rb                                                               
      &$5s_{1/2}$ & $0.0$     & $0.0$ & $0.0$         & $0.0$    &$0.0$    \\
      &$5p_{1/2}$ & $0.05756$ & $0.05754$ & $0.05758$ & $0.05759$&$0.05759$\\
      &$5p_{3/2}$ & $0.05869$ & $0.05867$ & $0.05871$ & $0.05872$&$0.05871$\\
      &$4d_{3/2}$ & $0.08854$ & $0.08854$ & $0.08834$ & $0.08836$&$0.08836$\\
      &$4d_{5/2}$ & $0.08854$ & $0.08853$ & $0.08834$ & $0.08836$&$0.08836$\\
      &&&&&\\                
$^{87}$Sr$^+$                                                   
      &$5s_{1/2}$ & $0.0$     & $0.0$ & $0.0$         & $0.0$    &$0.0$    \\
      &$4d_{3/2}$ & $0.06772$ & $0.06767$ & $0.06611$ & $0.06611$&$0.06607$\\      
      &$4d_{5/2}$ & $0.06882$ & $0.06877$ & $0.06724$ & $0.06724$&$0.06720$\\
      &$5p_{1/2}$ & $0.10836$ & $0.10837$ & $0.10840$ & $0.10841$&$0.10841$\\
      &$5p_{3/2}$ & $0.11217$ & $0.11217$ & $0.11220$ & $0.11221$&$0.11221$\\
\end{tabular}
\end{ruledtabular}
\end{table}


\subsection{Uncertainty estimates}

   Atomic properties calculated from RCC theory, in general, have three 
important sources of uncertanties. These are: omission of higher-$l$ orbital 
basis states, truncation of the dressed HFS operator $\tilde{H}_{\rm hfs} $ and
truncation of the coupled cluster operator $T $. The error arising from
the third source--is truncation of the CC operator--is almost mitigated with 
the inclusion of perturbative triples. So, effectively, HFS results presented 
in the Table. \ref{tab-hfs} have uncertainties from the first two sources. 
In the following we analyze and estimate the upper bound on the uncertanties
arising from each of these sources.

   The results presented in Table. \ref{tab-hfs} are the converged results
with the orbitals up to $h$ symmetry. 
To define the converged basis set, we 
do a series of calculations where we start with a minimal basis size of 112 
orbitals consisting of (1-14)$s$, (2-14)$p$, (3-15)$d$, (4-16)$f$,  and 
(5-14)$g$. Where we have used the non-relativistic notations for compact
representations. The basis set is increased by adding two orbitals in each 
symmetry in the successive sets of calculations till the change in the 
excitation energies and HFS constants are below $10^{-4}$. The values from 
different orbital basis sets are given in Table. \ref{ee_sets} and 
\ref{hfs_sets}. The total number of orbitals in the converged results is 177, 
and symmetry wise it is  (1-19)$s$, (2-19)$p$, (3-20)$d$, (4-21)$f$, (5-19)$g$ 
and (6-15)$h$. The single particle energies considered
are $\approx 10^5$ Hartrees for $s$, $p$, $d$ and $f$ orbitals and, 
$\approx 10^3$ Hartree for $g$ and $h$. 

   To estimate the uncertanties from excluding higher symmetries, we include 
orbitals of $i$ symmetry and compute the HFS constants. 
For $S_{1/2}$, the largest contribution is in the case of Rb and is about
0.07\%. However, for the states $P_{1/2}$ and $P_{3/2}$ it is Sr$^+$ which
has large contribution from the $i$ symmetry. These are 0.03\% and 0.04\%
respectively for the states $P_{1/2}$ and $P_{3/2}$. Unlike the $S_{1/2}$, 
$P_{1/2}$ and $P_{3/2}$ states, the triples contributions for $D_{3/2}$ and
$D_{5/2}$ are in general large. These are 1.2\% for $D_{3/2}$ in the case of
Rb and 7.0\% for $D_{5/2}$ in the case of Sr$^+$.
In the present implementation of RCC theory, incorporating $j$ and higher 
symmetry basis states renders the basis set too large for computations.
However, a leading-order analysis is possible with MBPT and
we find the contribution from $j$ symmetry is negligible.
%
%
\begin{table}[t]                
\caption{The HFS constants calculated using RCC for different sets of
         basis functions. All values listed are in units of MHz.}
\label{hfs_sets}
\begin{ruledtabular}                                                 
\begin{tabular}{ccccccc}                                               
Atom & State & Set 1 & Set 2  & Set 3 & Set 4 & Set 5\\
\hline                                                                  
$^{85}$Rb                                                               
    &$5s_{1/2}$ & $1029.19$&$1027.78$&$1030.70$&$1031.00$&$1030.94$\\
    &$5p_{1/2}$ & $120.66$ &$120.59$ &$120.66$ &$120.71$ &$120.69$ \\
    &$5p_{3/2}$ & $24.47$  &$24.45$  &$24.48$  &$24.48$  &$24.48$  \\
    &$4d_{3/2}$ & $7.54$   &$7.52$   &$7.84$   &$7.84$   &$7.85$   \\
    &$4d_{5/2}$ & $-4.78$  &$-4.74$  &$-4.79$  &$-4.78$  &$-4.78$  \\
    &&&&&\\                
$^{87}$Sr$^+$                                                   
    &$5s_{1/2}$&$-1012.82$&$-1012.97$&$-1014.09$&$-1014.20$&$-1014.20$\\
    &$5p_{1/2}$&$-178.61$ &$-178.63$ &$-178.67$ &$-178.75$ &$-178.73$ \\
    &$5p_{3/2}$&$-35.25$  &$-35.25$  &$-35.28$  &$-35.28$  &$-35.28$  \\
    &$4d_{3/2}$&$-45.90$  &$-45.92$  &$-46.21$  &$-46.31$  &$-46.30$  \\      
    &$4d_{5/2}$&$2.085$    &$2.07$    &$1.73$    &$1.72$    &$1.71$    \\
\end{tabular}
\end{ruledtabular}
\end{table}                                                                  
%
%
%
%
%
   To estimate the uncertanties arising from the truncation of
dressed properties operator, we resort to our previous work \cite{mani-10}. 
Where we proposed an iterative scheme to account for the higher order terms
in the dressed properties operator $\tilde H_{\rm hfs}$. Using this scheme,
we computed HFS constants contributions from the RCC terms which have 
{\em loe} of zero and one. These are the categories of diagrams which
contribute most to the atomic properties using RCC. Contributions from the
{\em loe} zero are 0.009\%, 0.008\%, 0.02\%, 0.03\% and
1.5\%, respectively for $S_{1/2}$, $P_{1/2}$, $P_{3/2}$, $D_{3/2}$ 
and $D_{5/2}$ states of Sr$^+$. From the {\em loe} one, however, these are 
large 0.09\%, 0.24\%, 0.35\%, 0.22\% and 12.40\%. The {\em loe } two and 
higher are not considered as it involve higher-order terms in T which will
naturally have smaller contributions.

   In addition to the sources of errors discussed so far, the other sources
of errors are the QED corrections. However, this cannot be considered as the 
source of error of the many-body method. It is rather the form of interaction
considered in the calculations.

   To put an upper bound on the uncertainty in the HFS results, we
select and add the largest change from each of the sources. Which 
turn out approximately to be 0.2\%, 0.3\%, 0.4\% and 1.5\%
for the states $S_{1/2}$, $P_{1/2}$, $P_{3/2}$ and $D_{3/2}$, respectively.
For the state $D_{5/2}$, since there is large cancellations, a comprehensive 
analysis is necessary to arrive at a meaningful uncertainty estimate. 


\section{Conclusions}

  We derive and propose a general tensor representation of the triple cluster 
operator is symmetric and with proper phase the core and virtual orbital 
indices may be permuted. The permuation properties are derived from the 
rules of angular momentum diagrams. Although, we have discussed the 
triple cluster operator in the context of valence triples $S_3$, the same
definition applies to core triples $ T_3$. This may be explicitly 
demonstrated with a minor topological transformation of the cluster 
operator. Based on the generalized tensor representation of $S_3$, we 
derive the linearized RCC equation for CCSDT approximation.

  We have analysed the contributions from the triple excitation cluster
operators to the HFS constants in detail. For the analysis we identify two
groups of diagrams depending on the nature of contractions to generate
the triple cluster operator. In most of the cases the LO and NLO are 
identified as $ A^6_{\rm ct}$  and $ A^3_{\rm vt}$. These arise from the
triples through the perturbation of $T_2$ cluster operators.
From the results, for HFS constants 
computations with perturebed triple excitation cluster operators, it is 
sufficient to approximate 
\begin{equation}
   A = A^1_{\rm cs} + A^3_{\rm cs} + A^6_{\rm ct} + A^3_{\rm vs} + 
       A^5_{\rm vs} + A^3_{\rm vt}.
\end{equation}
Contributions from the remaining terms are negligible and can be excluded from
the computations.

 The total number of diagrams considered in the present calculations are
108. Number of diagrams, however, will decrease significantly with geniune
valence triple cluster operators $S_3$ as the separation into core perturbed
and virtual perturbed cluster is not applicable. 
  
In conclusion, for calculations with the Gaussian type and $V^{N-1}$ 
potential orbital basis, the contributions from the triple excitation cluster operators is 
is at the most 0.6\%. 


\begin{acknowledgments}
We wish to thank Siddharth, Sandeep,  and
Arko  for useful discussions. The results presented in the paper
are based on computations using the HPC cluster at Physical Research
Laboratory, Ahmedabad.
\end{acknowledgments}


\end{document}